\documentclass[oldversion]{aa}  

\usepackage{graphicx,amsmath}
\usepackage{color}
\usepackage{natbib}		%for A&A : reimplements the LaTeX \cite command
\usepackage{amssymb} 
\usepackage{url}
\usepackage{multirow}
\bibpunct{(}{)}{;}{a}{}{,} % to follow the A&A style

\begin{document}

   \title{Modeling magnesium escape from HD\,209458b atmosphere.}
   				   
   \author{
   V.~Bourrier\inst{1,2}\and
   A.~Lecavelier des Etangs\inst{1,2}\and
   A.~Vidal-Madjar\inst{1,2}
	}
   
\authorrunning{V.~Bourrier et al.}
\titlerunning{Modeling magnesium escape from HD\,209458b}

\offprints{V.B. (\email{bourrier@iap.fr})}

   \institute{
   CNRS, UMR 7095, 
   Institut d'astrophysique de Paris, 
   98$^{\rm bis}$ boulevard Arago, F-75014 Paris, France
   \and
   UPMC Univ. Paris 6, UMR 7095, 
   Institut d'Astrophysique de Paris, 
   98$^{\rm bis}$ boulevard Arago, F-75014 Paris, France
   }
   
   \date{} %Received ...; accepted ...}
 
  \abstract
{

Transit observations in the Mg\,{\sc i} line of HD\,209458b revealed signatures of neutral magnesium escaping the upper atmosphere of the planet, while no atmospheric absorption was found in the Mg\,{\sc ii} doublet. Here we present a 3D particle model of the dynamics of neutral and ionized magnesium populations, coupled with an analytical modeling of the atmosphere below the exobase. Theoretical Mg\,{\sc i} absorption line profiles are directly compared with the absorption observed in the blue wing of the line during the planet transit. Observations are well-fitted with an escape rate of neutral magnesium $\dot{M}_{\mathrm{Mg^{0}}}$=2.9$\stackrel{+0.5}{_{-0.9}}\times10^{7}$\,g\,s$^{-1}$, an exobase close to the Roche lobe (\mbox{$R_\mathrm{exo}$=3$\stackrel{+1.3}{_{-0.9}}$$\,R_\mathrm{p}$}, where $R_\mathrm{p}$ is the planet radius) and a planetary wind velocity at the exobase $v_{\mathrm{pl-wind}}$=25\,km\,s$^{-1}$. The observed velocities of the planet-escaping magnesium up to -60\,km\,s$^{-1}$ are well explained by radiation pressure acceleration, provided that UV-photoionization is compensated for by electron recombination up to $\sim13\,R_\mathrm{p}$. If the exobase properties are constrained to values given by theoretical models of the deeper atmosphere ($R_\mathrm{exo}$=2$\,R_\mathrm{p}$ and $v_{\mathrm{pl-wind}}$=10\,km\,s$^{-1}$), the best fit to the observations is found at a similar electron density and escape rate within 2$\sigma$. In all cases, the mean temperature of the atmosphere below the exobase must be higher than $\sim6100$\,K. Simulations predict a redward expansion of the absorption profile from the beginning to the end of the transit. The spatial and spectral structure of the extended atmosphere is the result of complex interactions between radiation pressure, planetary gravity, and self-shielding, and can be probed through the analysis of transit absorption profiles in the Mg\,{\sc i} line.

}


   \maketitle

\section{Introduction}
\label{intro} 

The hot-Jupiter HD\,209458b has been the source of many detections of atomic and molecular species over the years (see \citealt{VM2013} and references therein). Transit observations of this planet in the H\,{\sc i} Lyman-$\alpha$ line led to the first detection of atmospheric escape (e.g., \citealt{VM2003,VM2008}; \citealt{BJ2007,BJ2008}; \citealt{Ehrenreich2008}). Heavier elements were identified at high altitudes in the extended exosphere of the planet in the lines of O\,{\sc i}, C\,{\sc ii}, and Si\,{\sc iii} (\citealt{VM2004}; \citealt{Linsky2010}, \citealt{BJ_Hosseini2010}), Si\,{\sc iv} (\citealt{Schlawin2010}) and more recently Mg\,{\sc i} (\citealt{VM2013}), supporting the idea that its atmosphere is in a state of ``blow-off''. \\
A large range of models have been developed to characterize the structure of the upper atmosphere of close-in giant exoplanets and to explain the evaporation process, either from theory or from observations (see \citealt{Bourrier_lecav2013} and references therein). Here we use the 3D numerical model detailed in Bourrier \& Lecavelier (2013), revised to interpret the observed escape of magnesium from the atmosphere of HD\,209458b. Transit observations of the planet in the Mg\,{\sc i} and Mg\,{\sc ii} lines are described in Section~\ref{obs}. We present our new model in Section~\ref{model}, notably the analytical modeling of the atmosphere below the exobase and the description of the ionization and recombination mechanisms. In Section~\ref{dyn}, we analyze the dynamics of an escaping magnesium atom, and in Section~\ref{regimes}, we describe the ionization state of the gas around HD\,209458b. In Section~\ref{model results}, we compare simulated spectra with the observations and put constraints on the exobase properties as well as the physical conditions in the extended atmosphere of HD\,209458b. Predictions of spectro-temporal variations in the absorption profile are presented in Section \ref{tempvar}.\\

\section{Observations in the Mg\,{\sc i} and Mg\,{\sc ii} lines}
\label{obs}

\begin{figure}[]
\includegraphics[trim=1cm 10cm 2cm 2cm, clip=true,width=\columnwidth]{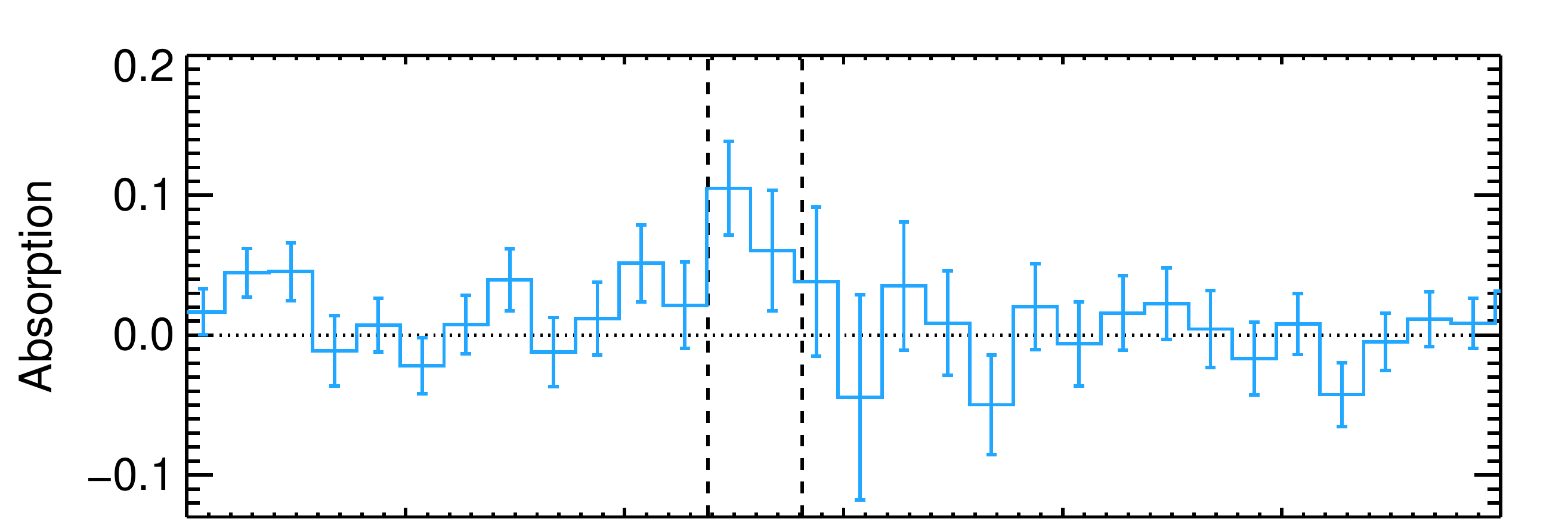}		
\includegraphics[trim=1cm 2cm 2cm 10.5cm, clip=true,width=\columnwidth]{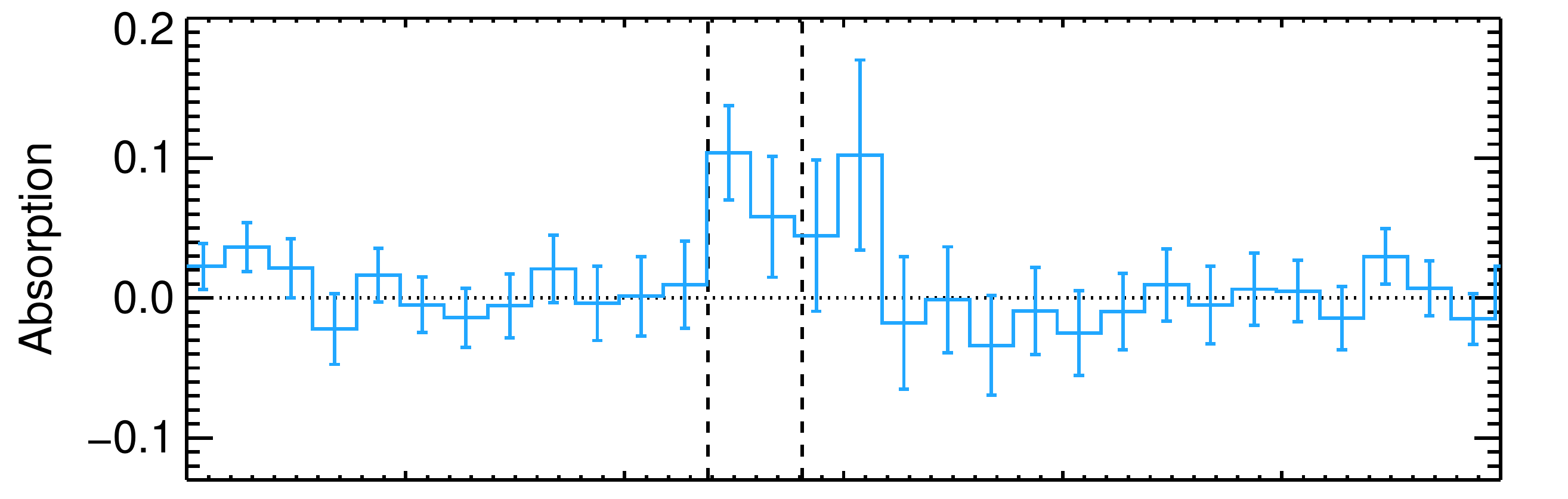}			
\includegraphics[trim=1cm 8cm 2cm 2.7cm, clip=true,width=\columnwidth]{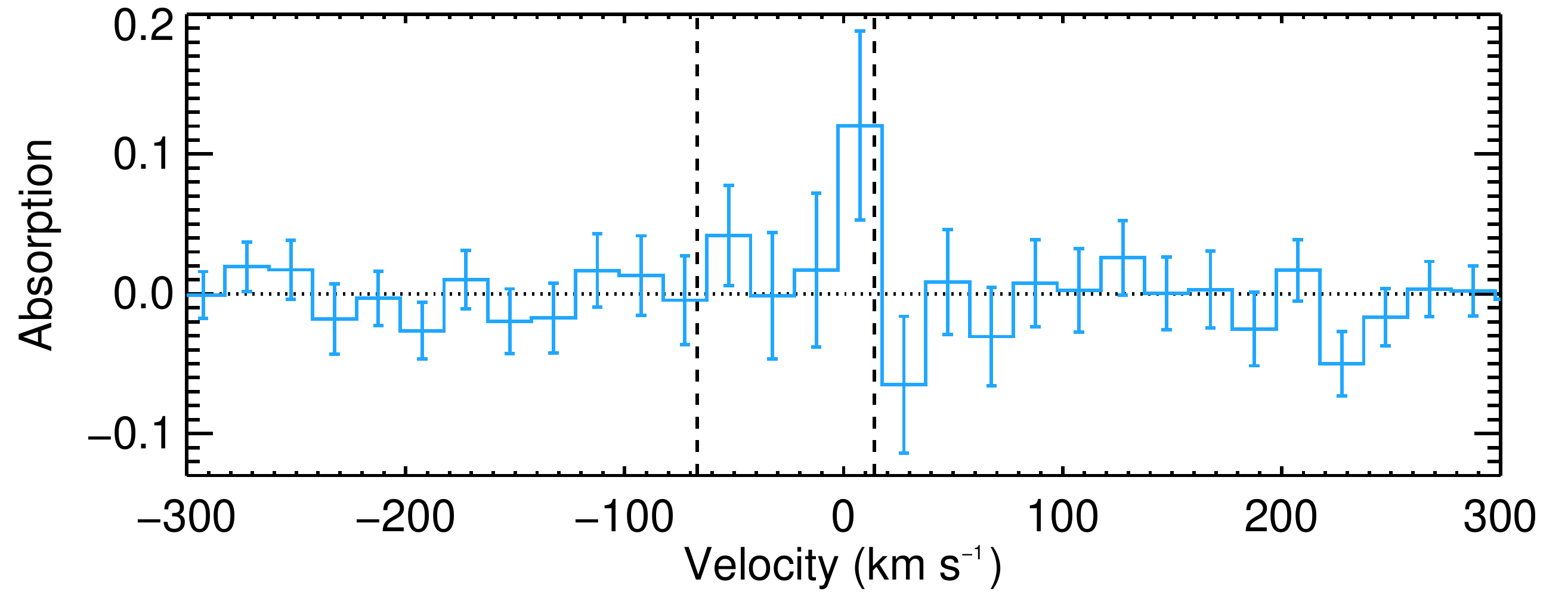}	
\caption[]{Mg\,{\sc i} line absorption profiles during the transit-ingress (top panel), transit-center (middle panel), and post-transit (lower panel) observations of HD\,209458b. The resolution is 20\,km\,$s^{-1}$. The absorption signatures detected by \citet{VM2013} are delimited by vertical black dashed lines.} 
\label{abs_3obs}
\end{figure}

\begin{figure}[]
\includegraphics[trim=1cm 10cm 2cm 2cm, clip=true,width=\columnwidth]{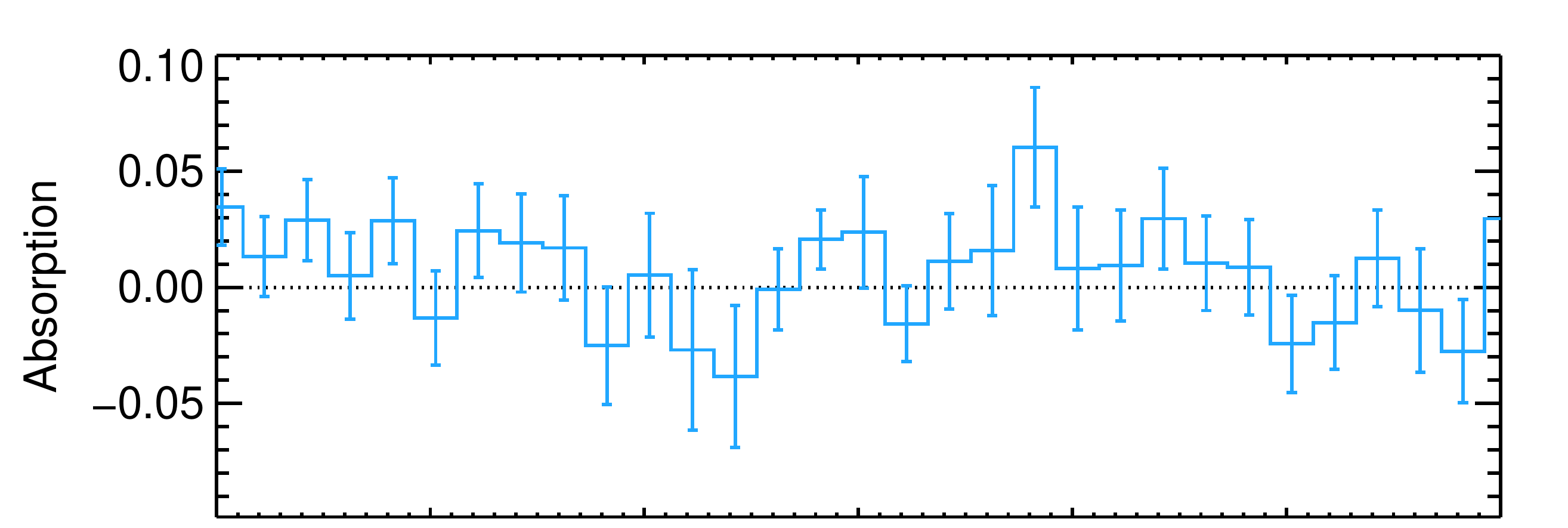}		
\includegraphics[trim=1cm 0.3cm 2cm 10.5cm, clip=true,width=\columnwidth]{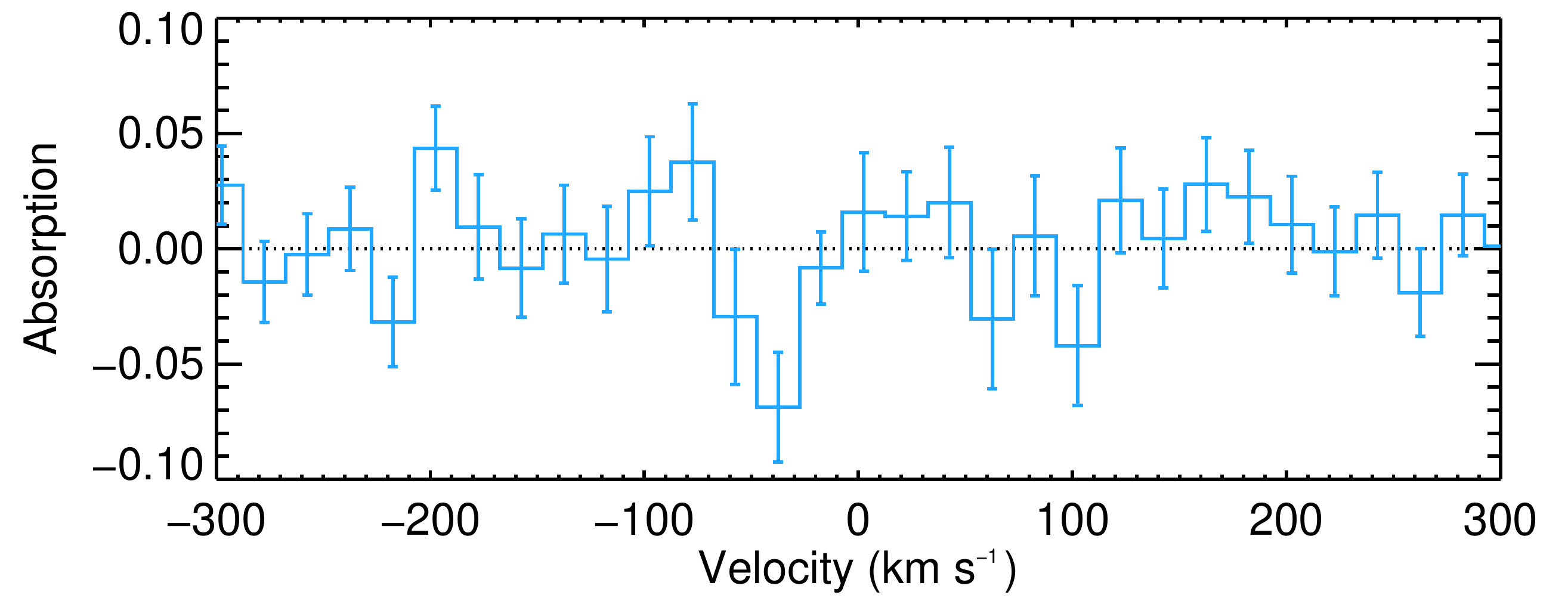}			
\caption[]{Absorption profiles in the doublet of singly ionized magnesium, calculated with the fluxes cumulated over the transit-ingress, transit-center, and post-transit observations. The resolution is 20\,km\,$s^{-1}$. No atmospheric absorption signature was detected in either the Mg\,{\sc ii}-k line (top panel) or the Mg\,{\sc ii}-h line (bottom panel).} 
\label{abs_3obs_ion}
\end{figure}

Three transits of HD\,209458b were observed by \citet{VM2013} in the Mg\,{\sc i} line of neutral magnesium (2852.9641\,\AA) and in the Mg\,{\sc ii} doublet of singly ionized magnesium (2796.3518 and 2803.5305\,\AA) in 2010 with the Space Telescope Imaging Spectrograph (STIS) instrument onboard the Hubble Space Telescope (HST). Each visit consists of two observations before the planetary transit, two observations during the transit labeled \textit{transit-ingress} and \textit{transit-center}, and one observation after the end of the transit labeled \textit{post-transit}. To get optimal signal-to-noise ratios \citet{VM2013} cumulated the spectra of the three visits. The time windows of the cumulated observations are given in Table~\ref{obs_log}. They discarded the first observation of each visit because of systematic trends known to affect STIS observations (e.g., \citealt{Brown2001} and \citealt{Charbonneau2002}). Here we will not use post-transit observations as a reference to calculate the absorption because of the possibility of an absorption signature related to the transit of a cometary tail after the planetary transit. In our study, we calculated all absorption depths in the same way by using only the second observation before the transit, labeled \textit{pre-transit}, as reference,
\begin{equation}
\label{eq:absorption}
A(\lambda)= 1- \frac{F(\lambda)}{F_{\mathrm{ref}(\lambda)}}.     			
\end{equation}
with $A(\lambda)$ the spectral absorption depth, $F$ the flux measured during the transit-ingress, transit-center, or post-transit observation, and $F_{\mathrm{ref}}$ measured during pre-transit observation. The corresponding absorption profiles are displayed in Fig.~\ref{abs_3obs} for the Mg\,{\sc i} line. \citet{VM2013} found a transit absorption signature in the blue wing of the Mg\,{\sc i} line, with a total depth of 8.8$\pm$2.1\% in the range -60 to -19\,km\,s$^{-1}$ if only the second observation before the transit is used as reference. A tentative detection of absorption was also reported in the post-transit observation with a depth of 4.0$\pm$2.2\% in the range -67 to 14\,km\,s$^{-1}$. No signal was found in excess of the planetary occultation depth in either line of the Mg\,{\sc ii} doublet (Fig.~\ref{abs_3obs_ion}). Hereafter, the reduced $\chi^2$ of the fits to the data are always found to be lower than 1 (e.g., Sect.~\ref{gen_approach}). Therefore, error bars on the spectral fluxes are likely over-estimated and, in the following, all error bars on estimated values of the model parameters can be considered as conservative.

\begin{table}[tbh]
\begin{tabular}{p{3cm}lcc}
\hline
\hline
\noalign{\smallskip}  
Cumulated    & \multicolumn{2}{c}{Time from center of transit}   \\		
data set     &	Start											&		End									 \\      
\noalign{\smallskip}
\hline
\noalign{\smallskip}
Pre-transit & -03:18  & -02:20  \\ 
\noalign{\smallskip}
Transit-ingress & -01:45  & -00:45  \\ 
\noalign{\smallskip}
Transit-center & -00:09   & \ 00:51    \\
\noalign{\smallskip}
Post-transit & \ 01:27  & \ 02:26    \\
\noalign{\smallskip}
\hline
\hline
\end{tabular}
\caption{Log of the observations of HD\,209458b used in this work, cumulated over three visits in 2010. Time is given in hours and minutes, and counted from the center of the transit. Start time always corresponds to the second visit (end time to the third visit) and for each observation the time windows of the three visits overlap.}
\label{obs_log}
\end{table}

\section{Model description}
\label{model}

\subsection{The case of magnesium}
\label{overview}

We used Monte-Carlo particle simulations to compute the dynamics of the populations of neutral and singly ionized magnesium in the extended atmosphere of HD\,209458b, and to estimate the physical conditions needed to reproduce the absorption profiles obtained by \citet{VM2013}. The general functioning of the model is described in Bourrier \& Lecavelier (2013), in which it was applied to the hydrogen escape from HD\,189733b and HD\,209458b. Improvements and modifications of the model needed to match the characteristics and specificities of magnesium are detailed hereafter. In the code, neutral and ionized magnesium atoms are represented by metaparticles made of the same number of atoms $N_{\mathrm{meta}}$. The total number of neutral magnesium particles launched every time step $dt$ depends on the \textit{escape rate of neutral magnesium} $\dot{M}_{\mathrm{Mg^{0}}}$. The particles are released from the entire upper atmosphere of the planet at the \textit{altitude of the exobase} $R_{\mathrm{exo}}$ (Sec.~\ref{thermo}). Hereafter, all numerical values of altitude will be given with the origin at the planet center. The initial velocity distribution of the particles, relative to the planet, is the combination of the \textit{radial bulk velocity of the planetary wind} $v_{\mathrm{pl-wind}}$ and an additional thermal component from a Maxwell-Boltzman velocity distribution corresponding to the temperature at $R_{\mathrm{exo}}$ (Sec.~\ref{ion-rec}). The modeling of the atmosphere below the exobase, characterized by its mean temperature $\overline{T}$, is described in Sect.~\ref{mainatm}.\\
The dynamics of neutral and ionized magnesium atoms above $R_{\mathrm{exo}}$ is constrained by the stellar radiation pressure (Sec.~\ref{rad_press}), the stellar gravity, and the planetary gravity. As the magnetic field of HD\,209458b is unknown (observations at radio wavelengths provided only an upper limit; \citealt{Lecav2011}), we neglected its influence on the charged particles' dynamics. Ionized magnesium particles may recombine into neutral magnesium through dielectronic recombination, which depends on the \textit{density of electron} $n_{\mathrm{e}}$ surrounding the planet. Ionized magnesium atoms are created by the stellar UV photoionization of neutral magnesium, and possibly through impacts with electrons (Sect.~\ref{ion-rec}).\\
Theoretical spectral absorption profiles are calculated in the Mg\,{\sc i} line and in both lines of the Mg\,{\sc ii} doublet via Eq 1., with simulated fluxes averaged over the time window of a given observation (transit-ingress, transit-center, or post-transit). The $\chi^2$ for the entire observation is the sum of the $\chi^2$ yielded by the comparison of each of the nine absorption profiles (i.e., three per line, as shown in Fig.~\ref{abs_3obs} for the Mg\,{\sc i} line) with the corresponding theoretical spectrum, calculated in the velocity range -300 to 300\,km\,s$^{-1}$. Because of the relatively small velocity range of the observed absorption signature and the high resolution of the STIS Echelle E230M grating spectra, the simulations have the same spectral resolution $\Delta v=$10\,km\,s$^{-1}$ as the observations.\\

To sum up, the free parameters of the model are the escape rate of neutral magnesium ($\dot{M}_{\mathrm{Mg^{0}}}$, in g\,s$^{-1}$), the electron density at a reference altitude of 3 planetary radii (hereafter called the reference electron density $n_{\mathrm{e}}(3R_{\mathrm{p}})$, in cm$^{-3}$), the altitude of the exobase ($R_{\mathrm{exo}}$, in $R_{\mathrm{p}}$), and the velocity of the planetary wind at the exobase ($v_{\mathrm{pl-wind}}$, in km\,s$^{-1}$). We also considered the influence of the main atmosphere temperature $\overline{T}$. We checked that simulations calculated with higher temporal and spectral resolutions produce the same theoretical absorption profiles. Physical parameters used in the model as well as numerical parameters with constant values are given in Table~\ref{num_param}.

\begin{table}[tbh]										
\begin{tabular}{llcccc}
\hline
\hline
\noalign{\smallskip}
Parameters 					  & Symbol  													   & Value       																							\\
\noalign{\smallskip}
\hline
\noalign{\smallskip}
Distance from Earth		& $D_{\mathrm{*}}$								 	 	 &    47.0\,pc        																		\\
\noalign{\smallskip}
Star radius						& $R_{\mathrm{*}}$								 		 &    1.146$\,R_{\mathrm{Sun}}$        	  							\\
\noalign{\smallskip}
Star mass							& $M_{\mathrm{*}}$								 		 &    1.148$\,M_{\mathrm{Sun}}$        										\\
\noalign{\smallskip}
Planet radius					& $R_{\mathrm{p}}$										 &    1.380$\,R_{\mathrm{Jup}}$       									\\
\noalign{\smallskip}
Planet mass						& $M_{\mathrm{p}}$								 		 &    0.69$\,M_{\mathrm{Sun}}$       											\\
\noalign{\smallskip}
Orbital period				& $T_{\mathrm{p}}$										 &    3.52475$\,days$        				    										\\
\noalign{\smallskip}
Semi-major axis				& $a_{\mathrm{p}}$								 		 &    0.047$\,AU$        																	\\
\noalign{\smallskip}
Inclination					  & $i_{\mathrm{p}}$								 		 &    $86.59^{\circ}$        															\\
\noalign{\smallskip}
\hline
\noalign{\smallskip}
Temporal resolution		&  $dt$																 &  $\approx$200\,s   																				\\
\noalign{\smallskip}
Spectral resolution 	& $\Delta v$     											 &  10\,km\,s$^{-1}$ 																				\\
\noalign{\smallskip}
Number of atoms       & \multirow{2}{*}{$N_{\mathrm{meta}}$} & \multirow{2}{*}{$1.02\times10^{30}$}   \\
per meta-particle     &																			 &																					  																		\\	
\noalign{\smallskip}
Mean temperature of   & \multirow{2}{*}{$\overline{T}$} & \multirow{2}{*}{$7000\,K$}   \\
the main atmosphere 	&    											            &  															\\	
\noalign{\smallskip}
\hline
\hline
\end{tabular}
\caption{Physical parameters for HD\,209458b, and numerical parameters with fixed values for all simulations.}
\label{num_param}
\end{table}

%%%%%%%%%%%%%%%%%%%%%%%%%%%%%%%%%%%%%%%%%%%%%%%%%%%%%%%%%%%%%%%%%%%%%%%

\subsection{Radiation pressure}
\label{rad_press}

\subsubsection{Radiation pressure from the Mg\,{\sc i}  line}

\begin{figure}[]
\includegraphics[trim=1cm 2cm 2cm 2cm, clip=true,width=\columnwidth]{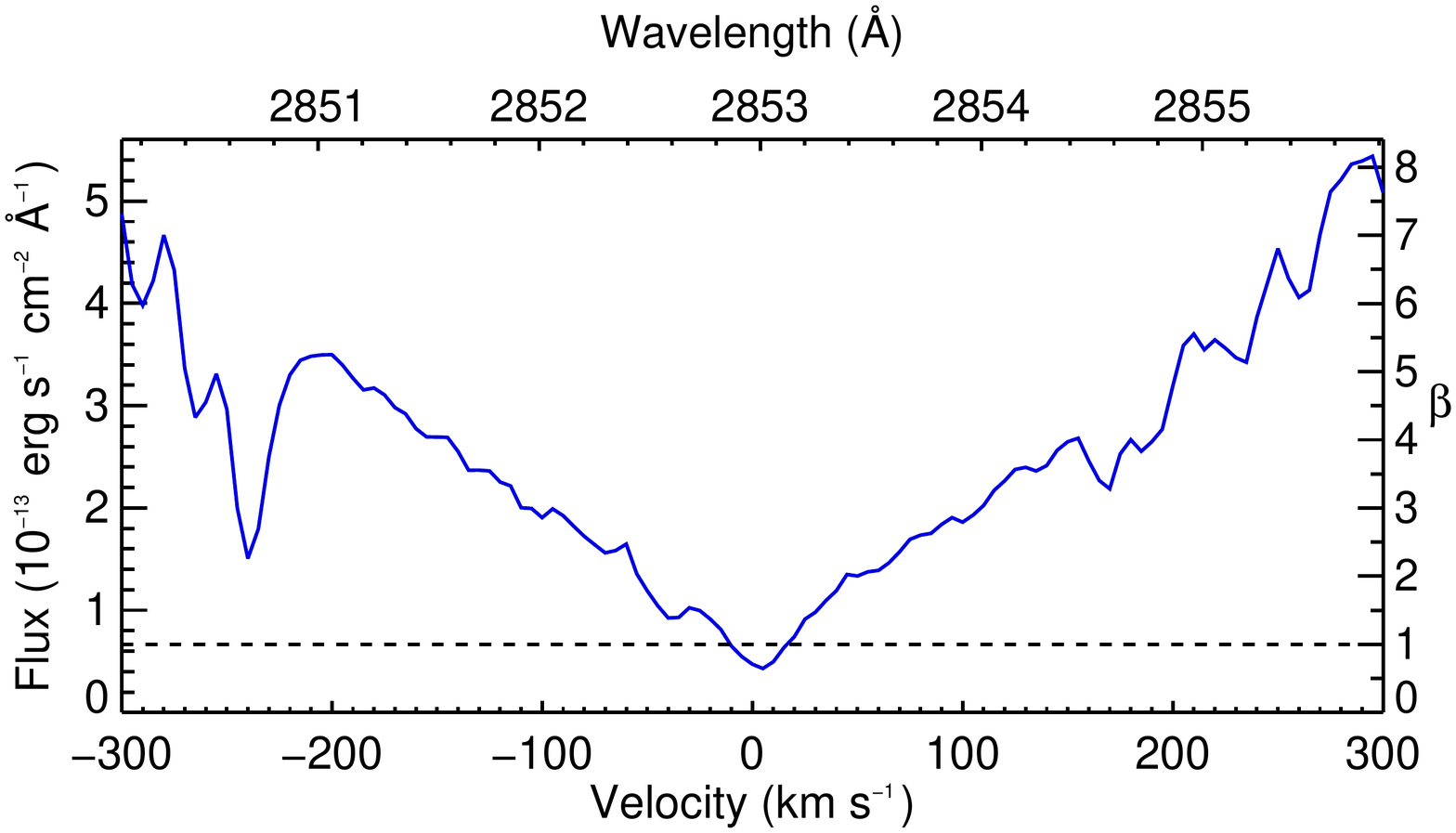}
\caption[]{Intrinsic Mg\,{\sc i} stellar absorption line profile of HD\,209458 as observed with HST/STIS at 2852.9641\,\AA\,(\citealt{VM2013}). The ratio $\beta$ between radiation pressure and stellar gravity (right axis) is proportional to the line flux. Note that $\beta$ is larger than 1 (horizontal dotted line) for velocities outside the range -10 to 15\,km\,s$^{-1}$, and increases steeply further away in the wings of the line.} 
\label{fit_HD209}
\end{figure}

A neutral magnesium atom in its ground state may be subjected to radiation pressure from Mg\,{\sc i} stellar lines at several wavelengths (e.g., at $\lambda$=2026.4768, 2852.9641, or 4572.3767\,\AA\,). However, only the Mg\,{\sc i} line at $\lambda_{\mathrm{0}}$=2852.9641\,\AA\,, characterized by a high oscillator strength (Table~\ref{dampingval}), has enough flux to generate a significant force on neutral magnesium atoms. We display the spectrum of this line in Fig.~\ref{fit_HD209}. All velocities herein are given in the reference frame of the star, which has a heliocentric radial velocity $V_{star}$=-14.7\,km\,s$^{-1}$ (\citealt{Kang2011}; \citealt{VM2013}). To obtain the radiation pressure applied to a neutral magnesium atom, we calculated the ratio $\beta$ between radiation pressure and stellar gravity. This coefficient is proportional to the stellar Mg\,{\sc i} line flux received by an escaping atom at its radial velocity (\citealt{Lagrange1998}; Fig.~\ref{fit_HD209}). In contrast to the H\,{\sc i} Lyman-$\alpha$ line (e.g., Bourrier \& Lecavelier 2013), whose central part absorbed by the interstellar medium and contaminated by geocoronal emission needs to be reconstructed, here observations can be used directly. As the flux in the Mg\,{\sc i} resonance line increases away from the line center, radiation pressure may even surpass stellar gravity when the radial velocity exceeds $\sim$10\,km\,s$^{-1}$ away from the star or $\sim$15\,km\,s$^{-1}$ toward the star, in which case atoms are more accelerated as their absolute radial velocity increases. Because radiation pressure has little effect on neutral magnesium particles at low velocities, planetary gravity may have significant effect on the atmospheric escape (see Sect.~\ref{exo_prop} and Sect.~\ref{gravSS}).

%%%%%%%%%%%%%%%%%%%%%%%%%%%%%%%%%%%%%%%%%%%%%%%%%%%%%%%%%%%%%%%%%%%%%%%
\subsubsection{Radiation pressure from the Mg\,{\sc ii} line}

\begin{figure}[]
\includegraphics[trim=2cm 2cm 2cm 2cm, clip=true,width=\columnwidth]{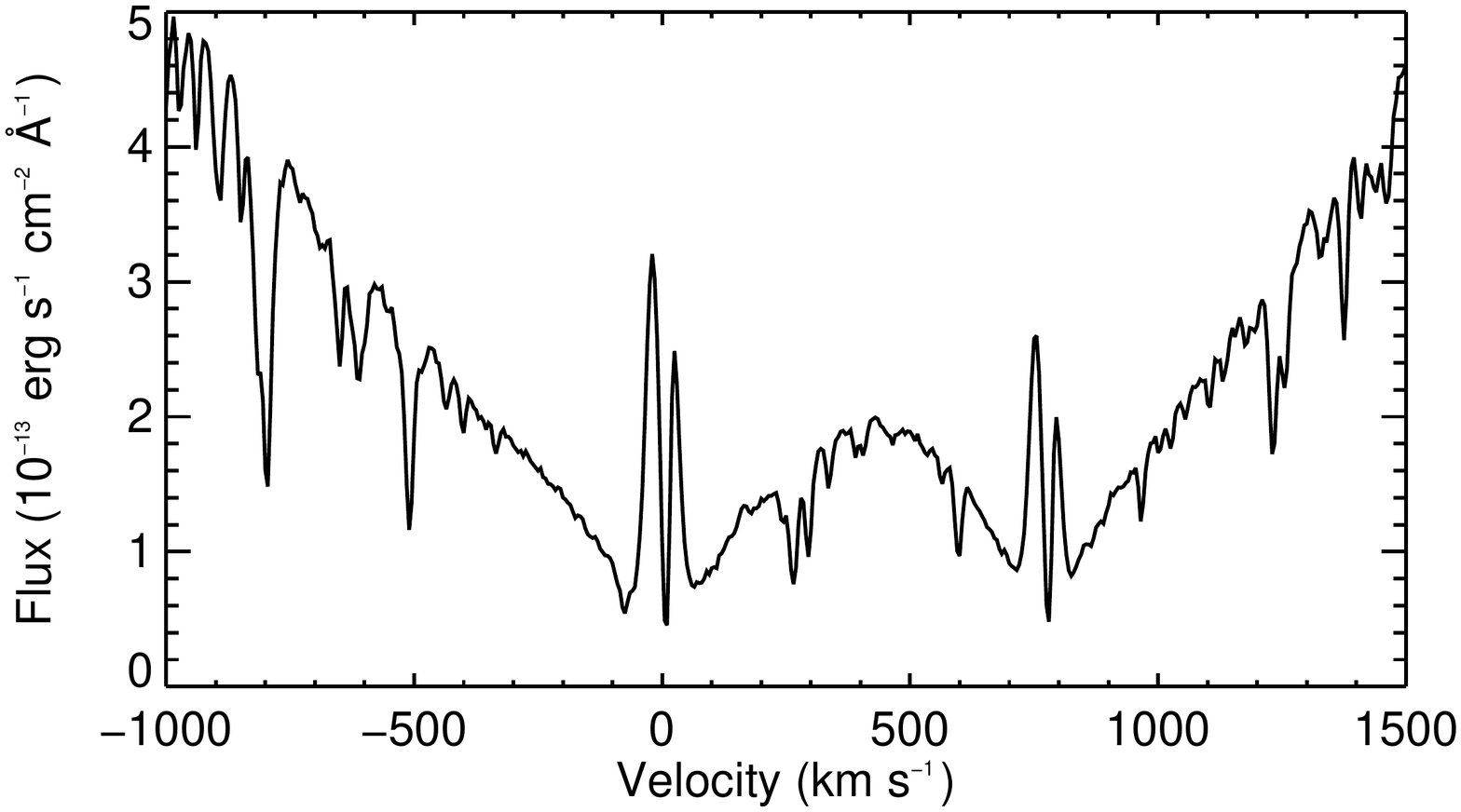}
\caption[]{Intrinsic Mg\,{\sc ii} stellar absorption line profile of HD\,209458 as observed with HST/STIS at 2796.3518\,\AA\, (Mg\,{\sc ii}-k line) and 2803.5305\,\AA\, (Mg\,{\sc ii}-h line). Velocities are relative to the Mg\,{\sc ii}-k line. Similarly to the Mg\,{\sc i} line, the Mg\,{\sc ii} lines spectral region is V-shaped.}
\label{full_mgII}
\end{figure}

\begin{figure}[]
\includegraphics[trim=1cm 1.5cm 2cm 2cm, clip=true,width=\columnwidth]{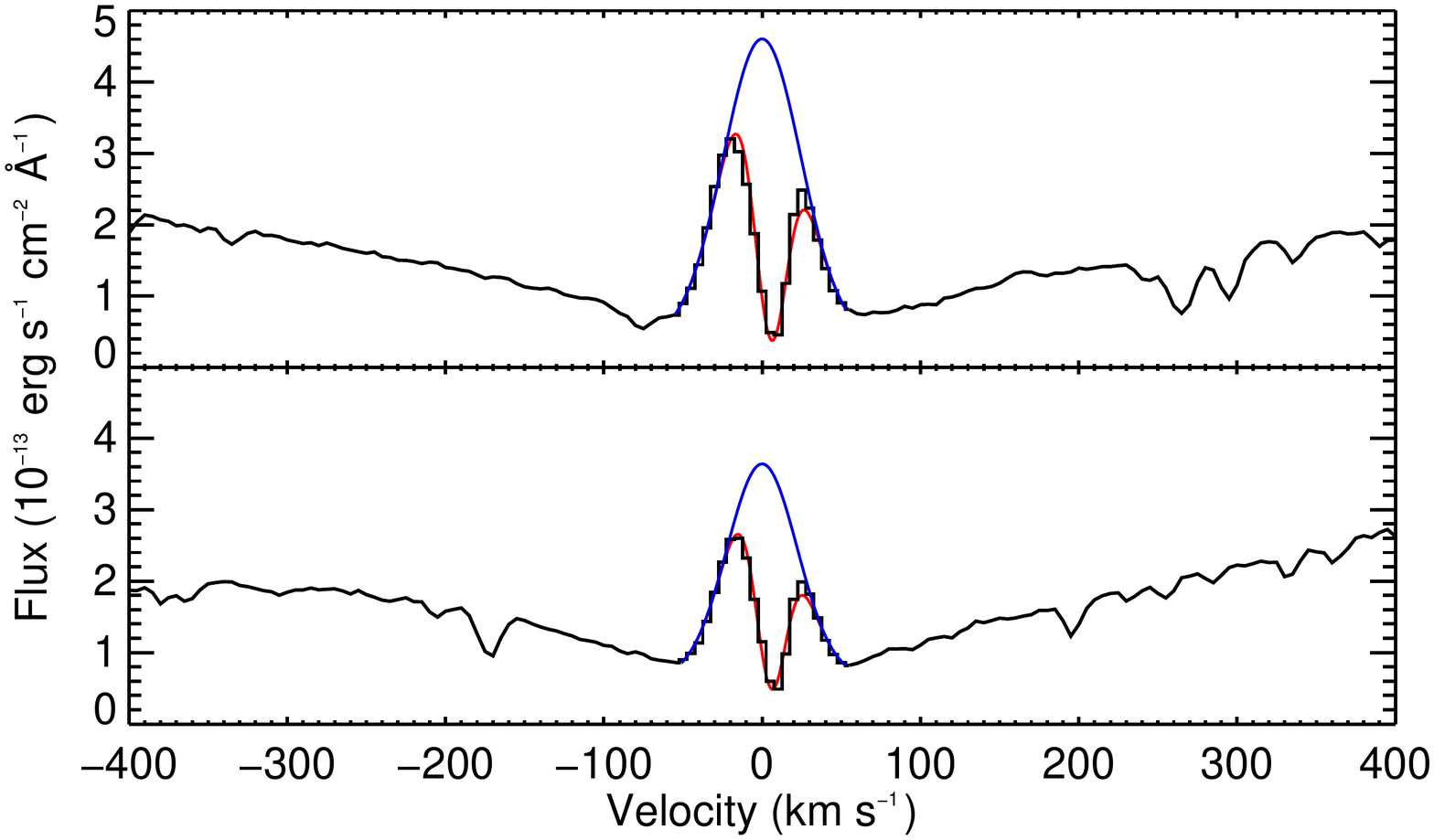}
\caption[]{Mg\,{\sc ii} stellar absorption line profile of HD\,209458 at 2796.3518\,\AA\, (Mg\,{\sc ii}-k line; top panel) and 2803.5305\,\AA\, (Mg\,{\sc ii}-h line; bottom panel). The blue line shows the theoretical self-reversed emission profile as seen by ionized magnesium atoms around the planet. The red line shows the line core profile after absorption by the interstellar Mg$^{+}$, to be compared with the observations (black histogram). Outside of the line core, observations (black line) can be used directly to calculate radiation pressure.}
\label{fit_HD209_MgII}
\end{figure}

\begin{figure}[]
\includegraphics[trim=1cm 3cm 2cm 2cm, clip=true,width=\columnwidth]{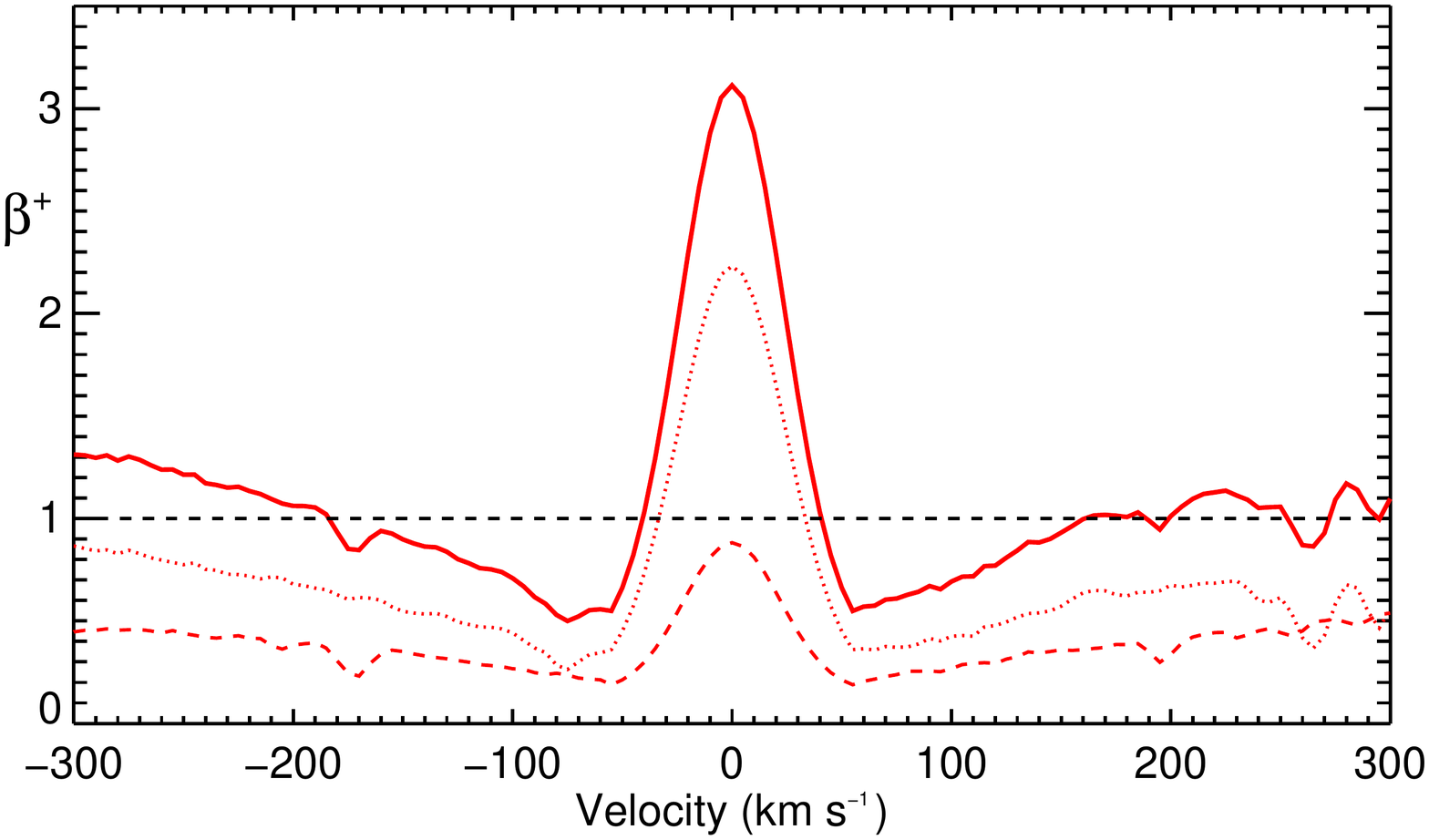}
\caption[]{Radiation pressure coefficients associated with the Mg\,{\sc ii}-k line (red dotted line) and the Mg\,{\sc ii}-h line (red dashed line), as a function of the Mg$^{+}$ ion radial velocity in the star reference frame. Their sum $\beta^{+}$ characterizes the total radiation pressure on an ionized magnesium particle (red solid line). $\beta^{+}$ is larger than 1 (horizontal black dotted line) in the line core and at high radial velocities in the wings of the line.} 
\label{beta_MgII}
\end{figure}

A singly ionized magnesium atom is subjected to radiation pressure from both lines of the Mg\,{\sc ii} doublet at 2796.3518\,\AA\, (Mg\,{\sc ii}-k) and 2803.5305\,\AA\, (Mg\,{\sc ii}-h), displayed in Fig.~\ref{full_mgII}. Observations can be used directly to calculate radiation pressure, except between about -55 and 55\,km\,s$^{-1}$ where stellar flux is absorbed by interstellar magnesium. In this range, a self-reversed chromospheric emission is observed, absorbed by the interstellar medium in a narrow band slightly redshifted with respect to the star velocity (Fig.~\ref{fit_HD209_MgII}). Using a similar procedure as in the case of the Lyman-$\alpha$ line (e.g., \citealt{Ehrenreich2011}), we fitted the line cores with two independent Gaussians centered on the star velocity, and modeled the interstellar medium (ISM) absorption with two Gaussians with different parameters except for a common center velocity. In the line shape, we did not consider the effect of the intrinsic self-reversal of the stellar lines, as they are shallow and limited to a narrow band in the line core (see the case of the Sun in \citealt{Lemaire1967}). In any case, this has no influence on our results as there is no significant amount of ionized magnesium at low altitudes in the atmosphere, where low-velocity particles are found (Sect.~\ref{model results}). \\
We replaced the doublet line cores in the observations by the reconstructed profiles. The radiation pressure coefficient from a given line is proportional to the flux received by a $Mg^{+}$ atom at its radial velocity with respect to the line center. The total radiation pressure coefficient $\beta^{+}$ is the sum of the coefficients from each line of the doublet (Fig.~\ref{beta_MgII}), with a stronger contribution from the Mg\,{\sc ii}-k line because of its higher flux and oscillator strength (Table~\ref{dampingval}). Ionized magnesium atoms are strongly accelerated by radiation pressure at low absolute radial velocities below $\sim$40\,km\,s$^{-1}$. Away from the chromospheric emission in the line core, the flux is lower and particles are decelerated until their velocity exceeds $\sim$185\,km\,s$^{-1}$ away from the star.

%%%%%%%%%%%%%%%%%%%%%%%%%%%%%%%%%%%%%%%%%%%%%%%%%%%%%%%%%%%%%%%%%%%%%%%
\subsection{Hydrodynamic and kinetic regimes}
\label{thermo}

\subsubsection{Exobase properties}
\label{exobase}

In contrast to hydrogen escape from HD\,209458b (\citealt{VM2004}) and HD\,189733b (\citealt{Lecav2012}), magnesium atoms are detected in a smaller velocity range close to the line center. At low velocities, escaping particles are likely near the planet (Bourrier \& Lecavelier 2013) and in the case of magnesium they can be significantly affected by planetary gravity. As a result, the initial launching conditions of magnesium particles ($R_{\mathrm{exo}}$ and $v_{\mathrm{pl-wind}}$) have a strong influence on the structure of the escaping cloud and the corresponding absorption profile.\\
Neutral magnesium particles are released at the exobase, i.e., the transition zone between the hydrodynamic regime and the kinetic regime. The different gases populating the atmosphere below the exobase (hereafter called the \textit{main atmosphere}) arrive in the transition zone in a state of ``blow-off'', as inferred from previous observations (\citealt{VM2004}; \citealt{Linsky2010} and references therein) and modeling of atmospheric escape from HD\,209458b (e.g., \citealt{Koskinen2013a}; \citealt{Guo2013}). The gases at the exobase are mixed in a global planetary wind with the velocity $v_{pl-wind}$. By contrast the kinetic regime corresponds to a collision-less region in which the dynamics of the gases can be described independently for each species. The precise altitude of the exobase of HD\,209458b is unknown, and we studied the impact of $R_{\mathrm{exo}}$ exploring values above 1.5\,$R_\mathrm{p}$. We also considered the influence of the planetary wind bulk velocity $v_{pl-wind}$ (Sect.~\ref{exo_prop}). The additional thermal speed component defined in Sect.~\ref{overview} is constrained by the temperature at the exobase level, given by the profile described in Sect.~\ref{ion-rec}.

%%%%%%%%%%%%%%%%%%%%%%%%%%%%%%%%%%%%%%%%%%%%%%%%%%%%%%%%%%%%%%%%%%%%%%%

\subsubsection{The atmosphere below the exobase}
\label{mainatm}

Atmospheric absorption in the core of the Mg\,{\sc i} line is partly due to the denser regions of the main atmosphere, which has to be modeled accurately. To calculate the optical depth and the self-shielding of the main atmosphere, we adopted an analytical depiction (as opposed to the particle modeling of the gas above the exobase) with a radial density profile corresponding to a hydrostatic equilibrium,

\begin{align}
\label{eq:hydroprof}
&n_{\mathrm{Mg^{0}}}(r)= n_{\mathrm{Mg^{0}}}(R_{\mathrm{exo}}) \, \exp\left( \frac{G \, M_{\mathrm{p}} \, \overline{m}  }{k_{\mathrm{B}} \, \overline{T} } ( \frac{1}{r} - \frac{1}{R_{\mathrm{exo}}} )\right) \\
&\text{for }r\,\leq\,{R_\mathrm{exo}}. \nonumber
\end{align}
This appears to be a good approximation of the density profile of the hydrodynamical atmosphere (e.g., \citealt{Koskinen2013b}). The main atmosphere extends up to $R_{\mathrm{exo}}$ and is characterized by its mean temperature $\overline{T}=$7000\,K (\citealt{Koskinen2013a}), its mean atomic weight $\overline{m} =$1.2 (assuming a Solar-like hydrogen-helium composition), and its density of neutral magnesium $n_{\mathrm{Mg^{0}}}(r)$ at the distance r from the planet center. In the numerical simulation, the density profile in the main atmosphere is reassessed by adjusting $n_{\mathrm{Mg^{0}}}(R_{\mathrm{exo}})$ to match the magnesium density obtained at the exobase for the particle calculation of the upper atmosphere in the stationary regime. During the transit of the main atmosphere, its surface is discretized with a high-resolution square grid (101$\times$101 cells). The column density of magnesium in front of each cell is given by
\begin{equation}
\label{eq:denscol}
N_{\mathrm{Mg^{0}}}(cell)=2 \, \int\limits_{\mathrm{0}}^{R_{\mathrm{exo}}}{\frac{n_{\mathrm{Mg^{0}}}(r) \, r}{\sqrt{r^2-r_{cell}^2}} \, dr},         
\end{equation} 
with $r_{cell}$ the distance between the planet center and its projection on the line of sight that goes through the cell.\\
We assumed the magnesium gas particles in the main atmosphere follow a Maxwellian velocity distribution. As a result, their absorption line profile is thermally broadened because of the gaussian component of the velocity distribution on the line of sight. It is characterized by its thermal width $v_\mathrm{th}=\sqrt{\frac{2\,k\,\overline{T}}{m_\mathrm{Mg^{0}}}}$ and its central velocity equal to the projection of the planet velocity on the line of sight. The profile width is further increased by natural Lorentzian broadening characterized by the damping constant $\Gamma$. The resulting absorption cross-section is thus a Voigt profile, which is multiplied by the column densities of the main atmosphere grid cells to obtain their optical depths. The addition of these values to the optical depth of the particles above the exobase yields the total optical depth of the neutral magnesium gas. The final absorption profile of the Mg\,{\sc i} stellar flux is obtained by combining the planetary occultation depth with neutral magnesium absorption, using Eq. (12) in Bourrier \& Lecavelier (2013). The absorption profile of the Mg\,{\sc ii} stellar flux is obtained in the same way but with the optical depth of ionized magnesium gas above the exobase only. We did not model ionization inside the main atmospere because no absorption was observed at low velocities in the Mg\,{\sc ii} line and observations are well reproduced with high electron densities, which ensure that magnesium remains neutral below the exobase (Sect.~\ref{model results}). See Table~\ref{dampingval} for constant values used in these calculations.

\begin{table}[tbh]
\begin{tabular}{cccc}
\hline
\hline
\noalign{\smallskip}
									&  Mg\,{\sc ii}-k & Mg\,{\sc ii}-h & Mg\,{\sc i}	   \\
\noalign{\smallskip}
\hline
\noalign{\smallskip}
$\lambda_{\mathrm{0}}$ (\,\AA)	& 2796.3518 &	2803.5305 & 2852.9641			\\
\noalign{\smallskip}
$f_{\mathrm{osc}}$ 						   & 0.615     & 0.306   & $1.83$       \\
\noalign{\smallskip}
$\Gamma$ (10$^{8}$\,s$^{-1}$)			& 2.62 & 2.59 	& 5.0 						\\
\noalign{\smallskip}
\hline
$m_{\mathrm{Mg}}$ (amu)								&  \multicolumn{3}{c}{$24.305$}	\\
\noalign{\smallskip}
\hline
\hline
\end{tabular}
\caption{Values of the constants used in the calculation of the radiation pressure exerted on Mg and Mg$^{+}$ particles, and their optical depth.}
\label{dampingval}
\end{table}

%%%%%%%%%%%%%%%%%%%%%%%%%%%%%%%%%%%%%%%%%%%%%%%%%%%%%%%%%%%%%%%%%%%%%%%

\subsection{Ionization and recombination mechanisms}
\label{ion-rec}

Neutral magnesium was observed up to -60\,km\,s$^{-1}$ in the blue wing of the Mg\,{\sc i} line. This velocity can be naturally explained by radiation pressure acceleration if the particles are accelerated over a sufficient period of time. However, magnesium atoms are exposed to UV photoionization from the star and at the short orbital distance of HD\,209458b they are quickly ionized with a timescale of about 0.6\,hours (\citealt{VM2013}). This time is too short for the particles to reach the observed velocities (Sect.~\ref{dyn}; Fig.~\ref{vel_profile}). \\
We used a scenario observed in the ISM (\citealt{Lallement1994}), in which neutral magnesium lifetime is extended through dielectronic recombination, compensating for ionization. With appropriate conditions for the electron density and temperature, ionized magnesium can indeed recombine with electrons into neutral magnesium. To calculate the transition probability of a neutral magnesium atom into an ionized atom, $dP_{\mathrm{0+}}$, we took into account the possibility that it is ionized sometime during $dt$ and then recombined with an electron. The reverse transition probability for an ionized magnesium atom $dP_{\mathrm{+0}}$ is calculated in the same way, and after integration we obtained
\begin{align}
\label{eq:proba}
dP_{\mathrm{0+}}=&\frac{\Gamma_{\mathrm{ion}}}{\Gamma_{\mathrm{ion}}-\Gamma_{\mathrm{rec}}}\left(exp(-dt\,\Gamma_{\mathrm{rec}})-exp(-dt\,\Gamma_{\mathrm{ion}})\right) \\				
dP_{\mathrm{+0}}=&\frac{\Gamma_{\mathrm{rec}}}{\Gamma_{\mathrm{rec}}-\Gamma_{\mathrm{ion}}}\left(exp(-dt\,\Gamma_{\mathrm{ion}})-exp(-dt\,\Gamma_{\mathrm{rec}})\right),  \nonumber	\end{align}\\
with $\Gamma_{\mathrm{ion}}$ the total ionization rate of neutral magnesium, and $\Gamma_{\mathrm{rec}}$ the total recombination rate of ionized magnesium. At the altitude $r$, $\Gamma_{\mathrm{rec}}$ depends on the recombination rate $\alpha_{\mathrm{rec}}(r)$ and the electron density $n_{\mathrm{e}}(r)$
\begin{equation}
\label{rate_rec}
\Gamma_{\mathrm{rec}}(r)=n_{\mathrm{e}}(r) \, \alpha_{\mathrm{rec}}(r). \\
\end{equation}\
In turn, $\alpha_{\mathrm{rec}}(r)$ (in cm$^{3}$\,s$^{-1}$) depends on the electron population temperature $T_{\mathrm{e}}(r)$ (in K) and was evaluated by \citet{Aldrovandi1973},
\begin{align}
\label{recrate}
\alpha_{\mathrm{rec}}(r)=&1.4\times10^{-13} \, \left(\frac{T_{\mathrm{e}}(r)}{10^{4}}\right)^{-0.855} \\
&+1.7\times10^{-3} \, T_{\mathrm{e}}(r)^{-\frac{3}{2}} \, e^{\frac{-5.1\times10^{4}}{T_{\mathrm{e}}(r)}}. \nonumber
\end{align}
The parameter $\Gamma_{\mathrm{ion}}$ depends on the constant photoionization rate $\Gamma_{\mathrm{UV-ion}}$ and, at the altitude $r$, on the electron-impact ionization rate $\alpha_{\mathrm{coll}}(r)$ and the electron density $n_{\mathrm{e}}(r)$,
\begin{equation}
\label{rate}
\Gamma_{\mathrm{ion}}(r)=\Gamma_{\mathrm{UV-ion}}+n_{\mathrm{e}}(r) \, \alpha_{\mathrm{coll}}(r). \\
\end{equation}
Without sufficient information about the densities of H$^{+}$ and He$^{+}$ in the atmosphere of HD\,209458b, we neglected the processes of charge-exchange of these particles with neutral magnesium. The UV flux of HD\,209458 was measured at the wavelengths below the ionization threshold (1621\,\AA\,; \citealt{VM2004}) and the photoionization rate is thus known to be $\Gamma_{\mathrm{UV-ion}}$=4.6\,s$^{-1}$ (\citealt{VM2013}). The ionization rate $\alpha_{\mathrm{coll}}(r)$ (in cm$^{3}$\,s$^{-1}$) was evaluated by \citet{Voronov1997},
\begin{align}
\label{eionrate}
\alpha_{\mathrm{coll}}(r)&=\frac{  0.621 \times 10^{-6} \times U(r)^{0.39} e^{-U(r)}    }{  0.592+U(r)     }    \\    
 U(r)&=\frac{ 88194.2 }{ T_{\mathrm{e}}(r) }. \nonumber
\end{align}
Because the electronic ionization rates calculated by \citet{Voronov1997} are for temperatures significantly above 11000\,K (at the upper limit of the temperature domain), and there are very large uncertainties on rates below this temperature, we did not take into account this mechanism, and we discuss its influence on our results in Sect.~\ref{conclu}. Hereafter, the ionization rate does not depend on the altitude and $\Gamma_{\mathrm{ion}}(r)=\Gamma_{\mathrm{UV-ion}}$.\\

\begin{figure}[]
\includegraphics[trim=1cm 2.5cm 2cm 2cm, clip=true,width=\columnwidth]{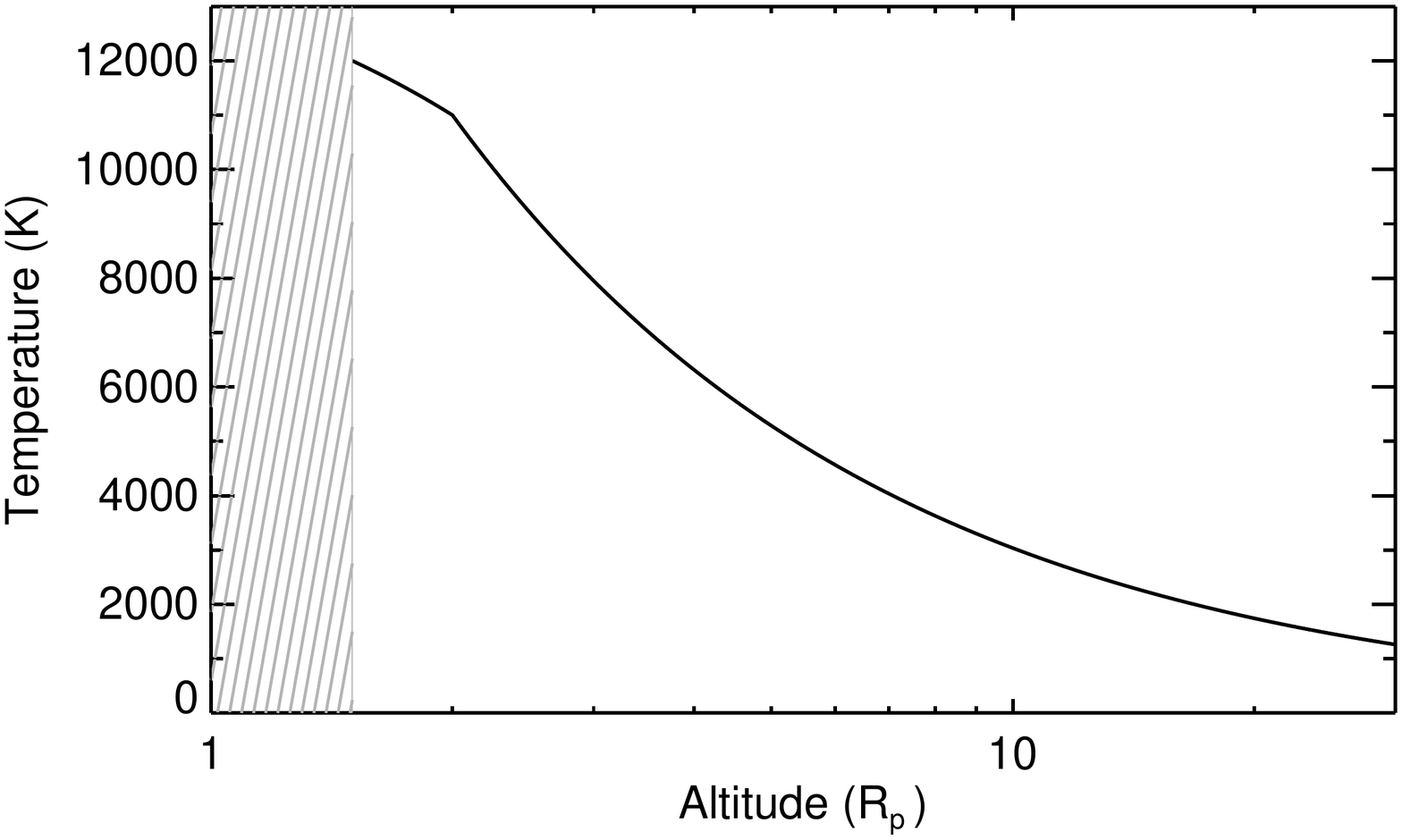}
\caption[]{Theoretical radial temperature profile $T_{\mathrm{e}}(r)$ of the electron population surrounding the planet as a function of the distance from the planet center. Profile was constructed by comparing different theoretical models and is defined for altitudes above 1.5\,${R_\mathrm{p}}$.}
\label{temp_profile}
\end{figure}

\begin{figure}[]
\includegraphics[trim=1cm 2cm 2cm 2cm, clip=true,width=\columnwidth]{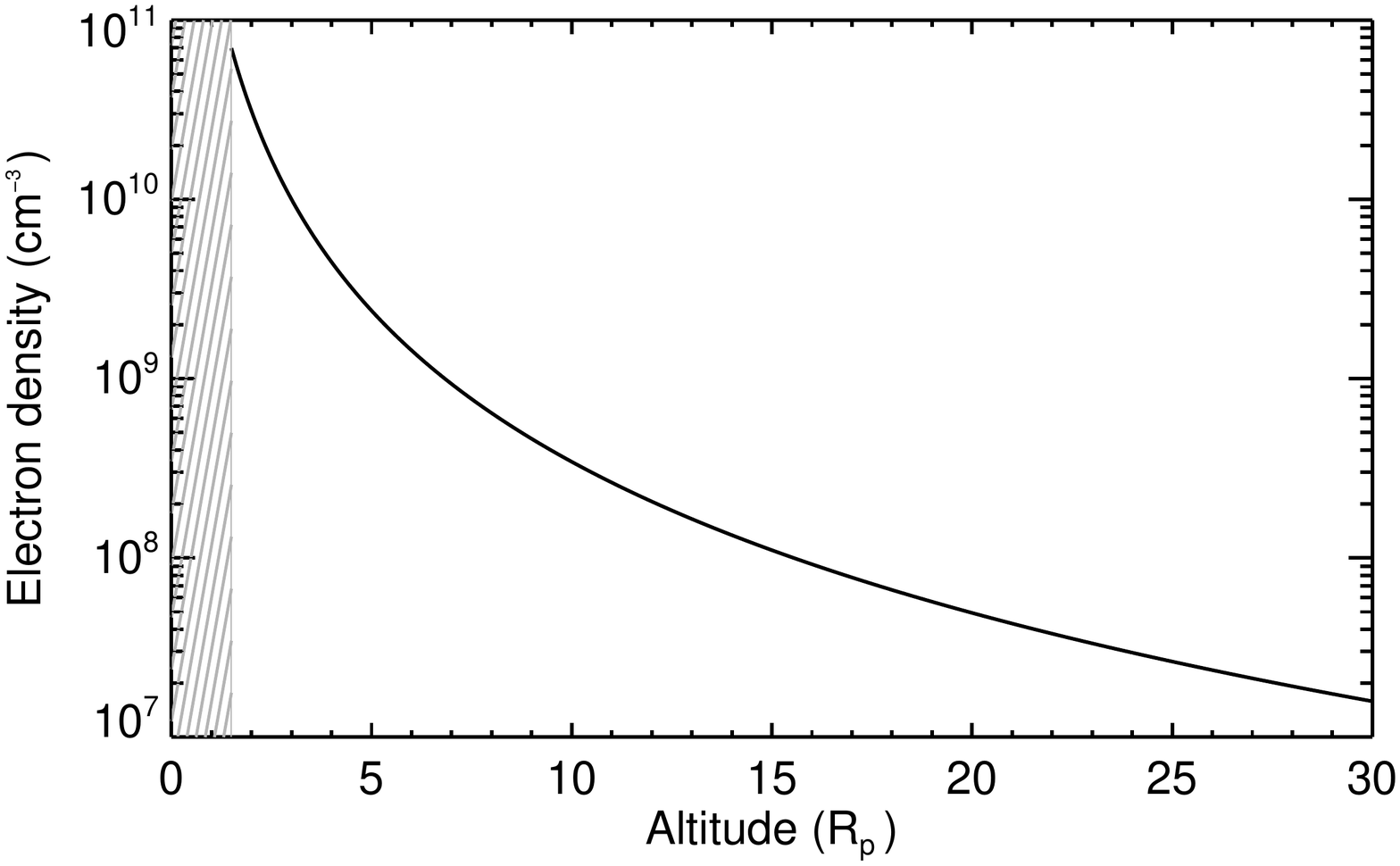}
\caption[]{Theoretical radial density profile $n_{\mathrm{e}}(r)$ of the electron population surrounding the planet as a function of the distance from the planet center, for $n_{\mathrm{e}}(3\,R_\mathrm{p})=10^{10}$\,cm$^{-3}$. Profile was constructed by comparing different theoretical models and is defined for altitudes above 1.5\,${R_\mathrm{p}}$.}
\label{ne_dens_profile_1e10}
\end{figure}

Several theoretical models have explored the temperature and density profiles in the upper atmosphere of HD\,209458b. We compared the works of \citet{GarciaMunoz2007}, \citet{Koskinen2013a}, and \citet{Guo2013} to infer an approximate radial temperature profile for the electron population (Fig.~\ref{temp_profile}),   
\begin{equation}
	\begin{array}{llll}
		T_{\mathrm{e}}=11000 \, \left(\frac{2\,R_\mathrm{p}}{r}\right)^{0.8} & K & \quad \text{for }   2&\leq\,\frac{r}{R_\mathrm{p}} \\
    T_{\mathrm{e}}=15000 - 2000 \, \left(\frac{r}{R_\mathrm{p}}\right)   & K & \quad \text{for } 1.5&\leq\,\frac{r}{R_\mathrm{p}}\,\leq2.			
    \end{array}
\label{tempprof}
\end{equation}
While the temperature in the atmosphere of HD\,209458b can be directly constrained by the observations, it is not the case with the electron density. From \citet{Koskinen2013a} and \citet{Guo2013} we estimated the shape of the radial density profile $n_{\mathrm{e}}(r)$, with the density at $r=3\,R_{\mathrm{p}}$ as a free parameter (Fig.~\ref{ne_dens_profile_1e10}),

\begin{equation}
\label{neprof}
n_{\mathrm{e}}(r)=n_{\mathrm{e}}(3\,R_\mathrm{p}) \, \left(\frac{3 \, R_\mathrm{p}}{r}\right)^{2.8}.	 \\
\end{equation}

\section{Magnesium atoms dynamics}
\label{dyn}

\begin{figure}[]
\includegraphics[trim=1cm 2.5cm 2cm 2cm, clip=true,width=\columnwidth]{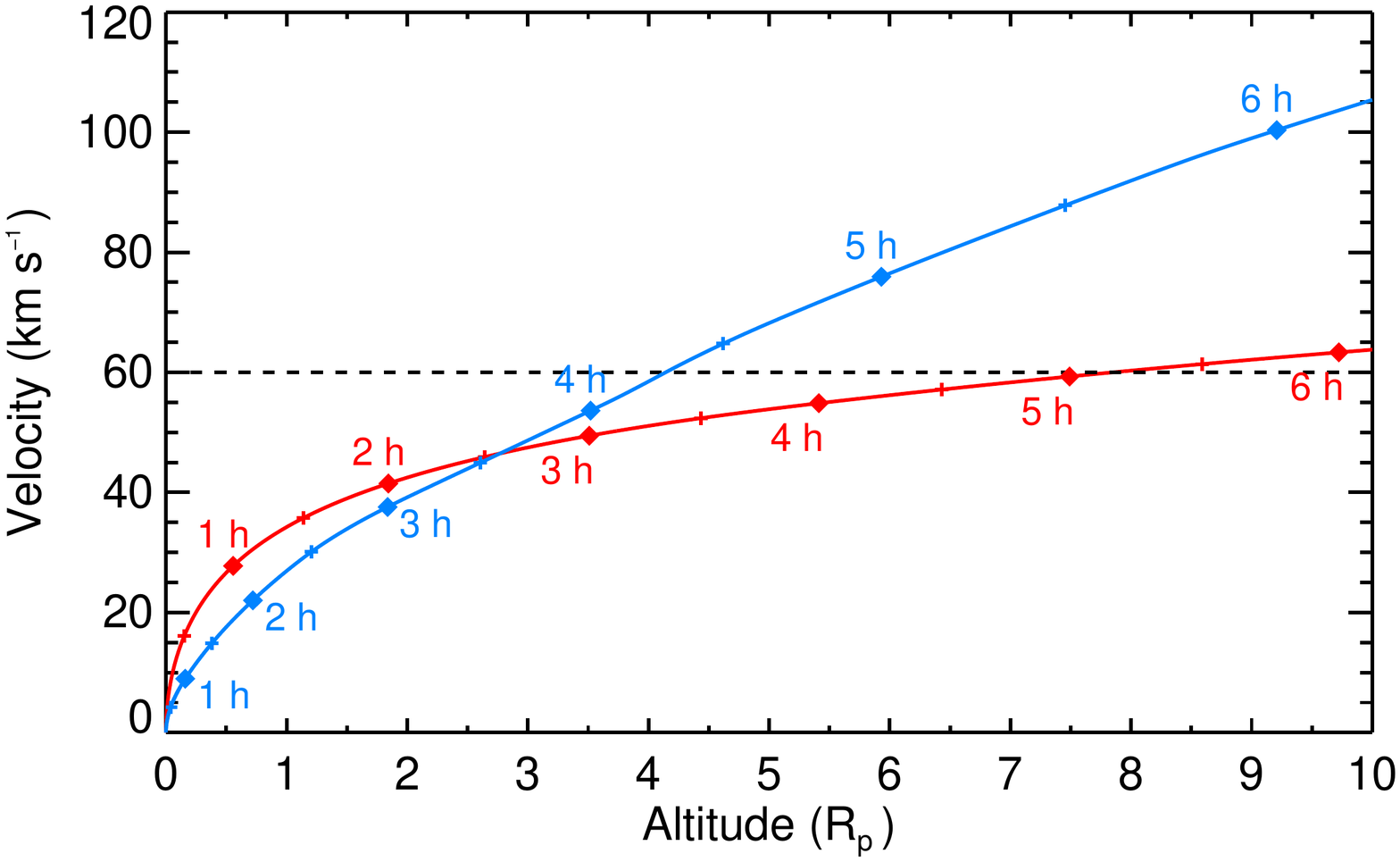}
\caption[]{Blueward radial velocity of a neutral (blue line) or ionized (red line) magnesium atom escaped from HD\,209458b, as a function of the distance from the planet center. Diamond symbols show the time since the particle escape with a step of one hour (half-hours are indicated by cross symbols). A neutral particle reach 60\,km\,s$^{-1}$ (horizontal dotted line) in about 4.25\,h at 4\,$R_{\mathrm{p}}$ and an ionized particle in about 5.2\,h at 8\,$R_{\mathrm{p}}$.}
\label{vel_profile}
\end{figure}

In a first approach to understanding the processes in play, we used a 2D toy model to estimate the velocity profiles of a neutral and an ionized magnesium atom escaping the planet atmosphere. The particle is launched with the planet orbital velocity in the stellar referential. Its trajectory in the orbital plane is then constrained only by stellar gravity and radiation pressure. Note that radiation pressure depends on the particle radial velocity relative to the star, which differs from its velocity as observed from the Earth (\citealt{Bourrier_lecav2013}). Fig.~\ref{vel_profile} displays the former as a function of the particle distance from HD\,209458b. Because the particle initial velocity is that of the planet, its radial velocity is null. In the case of a neutral magnesium atom, $\beta$ is thus lower than unity (Fig.~\ref{fit_HD209}), but the particle quickly separates from the planet orbit and gains enough radial velocity through the projection of the planet velocity on the star/particle axis to be accelerated away from the star by radiation pressure. The particle acceleration and velocity keep increasing as the stellar flux absorbed in the blue wing of the Mg\,{\sc i} line increases away from the line center. On the contrary an ionized magnesium atom is strongly accelerated at low velocity because of the chromospheric emission in the Mg\,{\sc ii} line core (Fig.~\ref{beta_MgII}). The particle radial velocity then keeps increasing, but at a much lower rate when $\beta^{+}$ is below unity in the wings of the line (Fig.~\ref{vel_profile}).\\
With a characteristic UV-photoionization lifetime of about 0.6\,hours, a magnesium atom has a low probability of remaining neutral until it is accelerated to the upper limit of the signature observed in the blue wing at -60\,km\,s$^{-1}$ (Fig.~\ref{vel_profile}), unless its lifetime is extended through electron recombination. A complete 3D modeling is needed to fully simulate the dynamics of the escaping magnesium gas, as the magnesium atoms will alternate between ionized and neutral states and their trajectory may be significantly altered by planetary gravity and self-shielding effects.

\section{Mg-recombining layer of the exosphere}
\label{regimes}

The amount of magnesium atoms in the neutral or the ionized state depends on the relative values of the UV-photoionization rate and the recombination rate. As will be confirmed by the detailed simulation (Sect.~\ref{model results}), in a first approach to describe the global picture of ionization/recombination processes we can neglect the particles's dynamics and consider that in the stationary regime there are the same number of ionizations and recombinations at the altitude $r$, i.e.,
\begin{equation}
N_\mathrm{Mg^{+}}(r)\,dP_{\mathrm{+0}}(r)=N_\mathrm{Mg^{0}}(r)\,dP_{\mathrm{0+}}.
\label{dnion_dnrec}
\end{equation}
Using Eq.~\ref{eq:proba}, we obtained an approximation of the ratio between the ionized and neutral magnesium populations at the altitude $r$
\begin{equation}
\frac{N_\mathrm{Mg^{0}}(r)}{N_\mathrm{Mg^{+}}(r)}=\frac{\Gamma_{\mathrm{rec}}(r)}{\Gamma_{\mathrm{UV-ion}}}.
\label{pop_ratio}
\end{equation}

\begin{figure}[]
\includegraphics[width=\columnwidth]{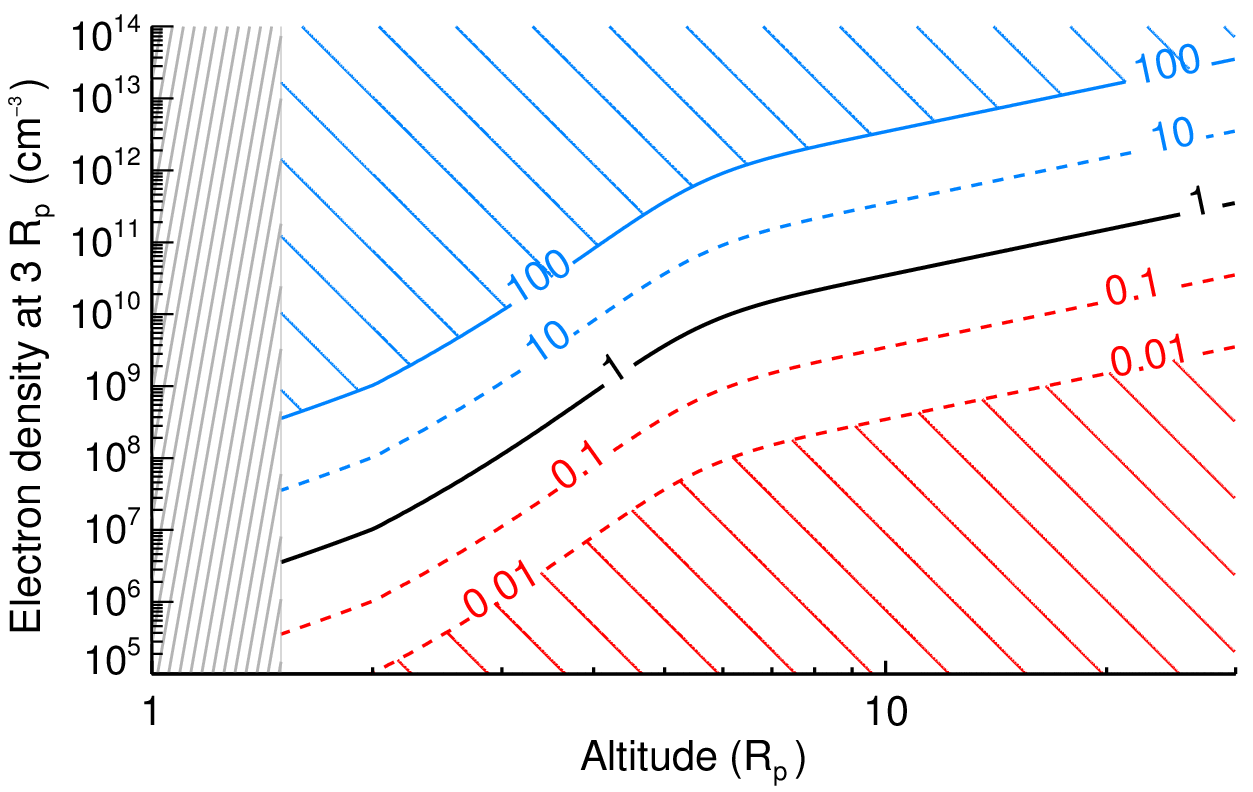}
\caption[]{Ratio between the recombination rate $\Gamma_{\mathrm{rec}}$ and the ionization rate $\Gamma_{\mathrm{UV-ion}}$, as a function of the altitude counted from the planet center and the electron density at 3\,$R_{\mathrm{p}}$. Solid and dashed lines show isovalues of the ratio. In the striped red zone photoionization is dominant ($\Gamma_{\mathrm{UV-ion}}\ge100\Gamma_{\mathrm{rec}}$) and the probability that an ionized particle recombines is close to zero. With altitudes below the equilibrium altitude (solid black line), the lifetime of a neutral magnesium atom is significantly extended, and in the blue striped zone below the recombination altitude (solid blue line) the probability that it is ionized is close to 0 ($\Gamma_{\mathrm{rec}}\ge100\Gamma_{\mathrm{UV-ion}}$).}
\label{rate_recAP}
\end{figure}

For a given electron density profile, i.e., a given value of $n_{\mathrm{e}}(3\,R_\mathrm{p})$, this ratio is only dependent on the distance from the planet (Fig.~\ref{rate_recAP}). If the reference electron density is lower than about $10^{5}$\,cm$^{-3}$ the UV photoionization mechanism is dominant at all altitudes above the exobase ($\Gamma_{\mathrm{rec}}/\Gamma_{\mathrm{UV-ion}}\le$0.01; red striped zone in Fig.~\ref{rate_recAP}). Neutral magnesium are quickly ionized with a short lifetime of about 0.6\,h and are unlikely to recombine ($dP_{\mathrm{+0}}(r)\sim0$). With higher reference electrons densities, there are enough electrons in a shell above the exobase to significantly extend the lifetime of escaping neutral magnesium atoms through recombination. We define the recombination altitude $R_{\mathrm{rec}}$ ($\Gamma_{\mathrm{rec}}/\Gamma_{\mathrm{UV-ion}}=$100), i.e., the critical altitude below which the probability that a magnesium atom remains neutral is close to unity (blue striped zone in Fig.~\ref{rate_recAP}). The recombination altitude is included in the \textit{Mg-recombining layer of the exosphere} (hereafter called ``Mg-rec layer''), which is defined as the layer where the recombination rate is always higher than the ionization rate. It extends from the exobase to the equilibrium altitude $R_\mathrm{eq}$ ($\Gamma_{\mathrm{rec}}=\Gamma_{\mathrm{UV-ion}}$). Within the Mg-rec layer, magnesium atoms have spectroscopic signatures in the Mg\,{\sc i} line, which also drives their radiation-pressure acceleration. Beyond the equilibrium altitude, ionized particles do not recombine efficiently and their dynamics and absorption signatures are driven by the Mg\,{\sc ii} doublet. The upper limit of the absorption profile in the blue wing of the Mg\,{\sc i} line thus increases with $R_\mathrm{eq}$, i.e., with the reference electron density. For example, for magnesium atoms to absorb in the Mg\,{\sc i} line up to 60\,km\,s$^{-1}$, they must remain mostly neutral up to about 4$R_{\mathrm{p}}$ (Fig.~\ref{vel_profile}), which means this altitude is somewhere between the recombination altitude ($n_{\mathrm{e}}(3\,R_\mathrm{p})$=10$^{11}$\,cm$^{-3}$) and the equilibrium altitude ($n_{\mathrm{e}}(3\,R_\mathrm{p})$=10$^{9}$\,cm$^{-3}$).

%%%%%%%%%%%%%%%%%%%%%%%%%%%%%%%%%%%%%%%%%%%%%%%%%%%%%%%%%%%%%%%%%%%%%%%

\section{Modeling magnesium escape from HD\,209458b}
\label{model results}

\subsection{General scenario}
\label{gen_approach}

\subsubsection{Best-fit parameters}
\label{abs_bestfit}

\begin{figure}[]
\includegraphics[trim=1cm 1.5cm 2cm 2cm, clip=true,width=\columnwidth]{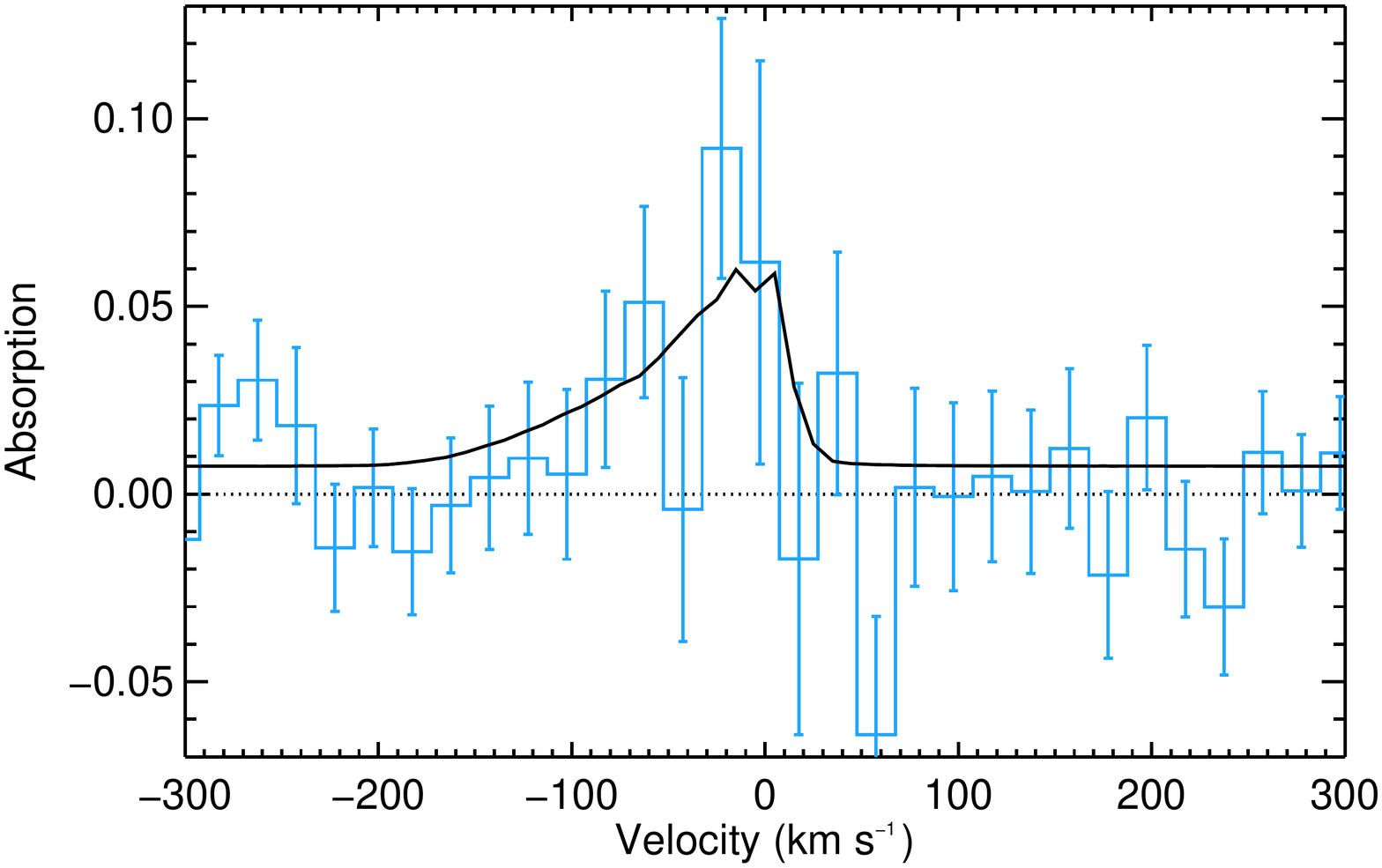}	 
\caption[]{Absorption profiles in the Mg\,{\sc i} line calculated with the fluxes averaged over transit and post-transit observations of HD\,209458b. The observed profile (\citealt{VM2013}) is displayed as a blue histogram with a resolution of 20\,km\,$s^{-1}$, while the solid black line corresponds to the theoretical global best-fit profile. The $\chi^2_{MgI}$ associated with this absorption profile is 225.9 (357 data points), and the $\chi^2$ for all observations is 802.4 for 1067 degrees of freedom.} 
\label{best_absprof}
\end{figure}

\begin{figure}[]
\includegraphics[trim=1cm 10cm 2cm 2cm, clip=true, clip=true,width=\columnwidth]{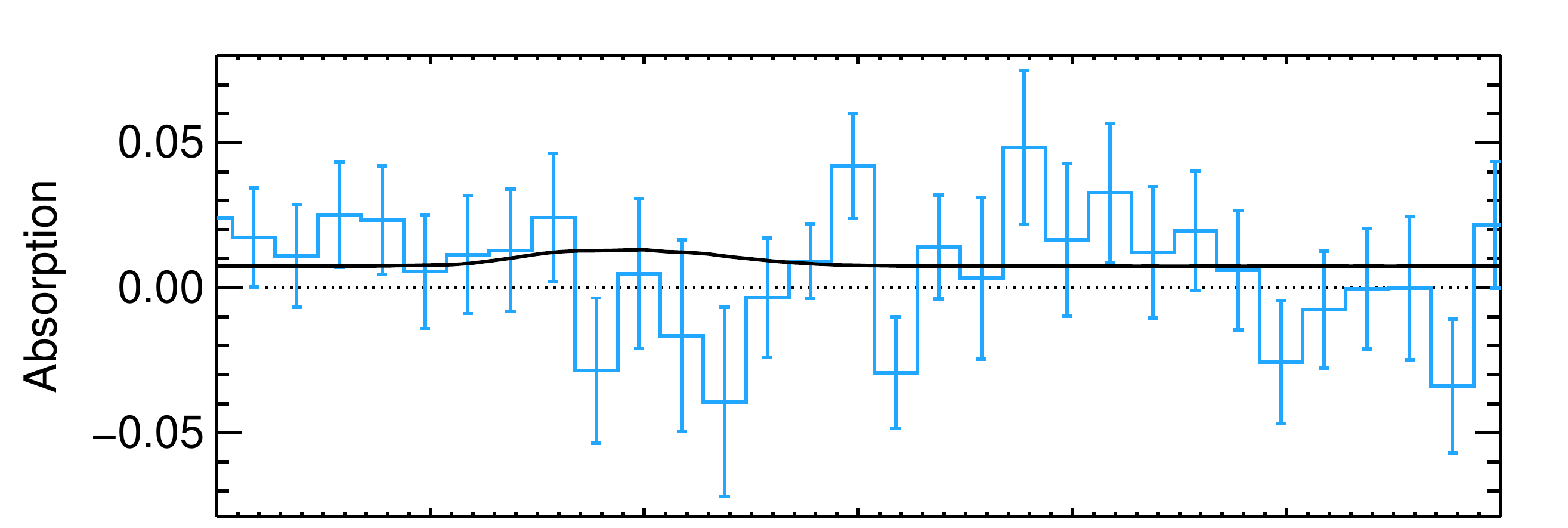}	
\includegraphics[trim=1cm 0cm 2cm 10.5cm, clip=true, clip=true,width=\columnwidth]{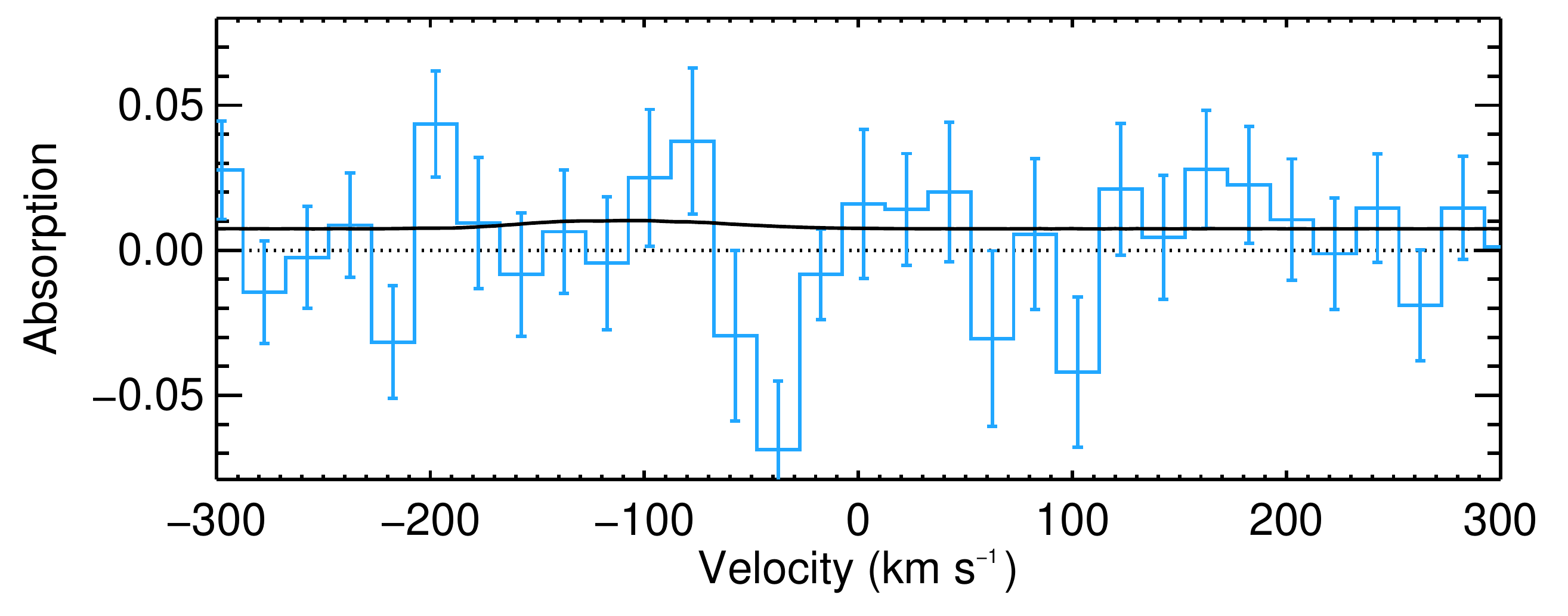}		
\caption[]{Same plot as in Fig.~\ref{best_absprof} with absorption profiles in the Mg\,{\sc ii}-k line (upper panel) and Mg\,{\sc ii}-h line (lower panel) for the global best-fit simulation. They are obtained with $\chi^2_{MgIIk}$=261.6 (357 data points) and $\chi^2_{MgIIh}$=314.9 (357 data points), and a total $\chi^2$=802.4 for 1067 degrees of freedom for all observations.}
\label{best_absprof_MgII}
\end{figure}

We used our 3D model to estimate the exobase properties and the physical conditions in the exosphere of HD\,209458b needed to reproduce the observations in the lines of neutral and singly ionized magnesium. In this general scenario (hereafter, G-scenario), all four parameters of the model were let free to vary. The best fit was obtained with a $\chi^2$ of 802.4 for 1067 degrees of freedom, when particles escape the atmosphere at $R_{\mathrm{exo}}=3\,R_{\mathrm{p}}$ (exobase level) with a planetary wind velocity $v_{\mathrm{pl-wind}}$=25\,km\,s$^{-1}$, the escape rate of neutral magnesium is $\dot{M}_{\mathrm{Mg^{0}}}$=2.9$\times10^{7}$\,g\,s$^{-1}$ and the reference electron density is $n_{\mathrm{e}}(3R_{\mathrm{p}})$=6.4$\times10^{10}$cm$^{-3}$. We caution that with $\chi^2$ lower than the number of degrees of freedom, error bars on these parameters are conservative values. Absorption profiles from this \textit{global} best-fit simulation are displayed in Fig.~\ref{best_absprof} for the Mg\,{\sc i} line and in Fig.~\ref{best_absprof_MgII} for the Mg\,{\sc ii} doublet. In the following sections, we study the structure of the escaping cloud in relation with the absorption it generates, and discuss the uncertainty on the derived values of the four model parameters.\\

%%%%%%%%%%%%%%%%%%%%%%%%%%%%%%%%%%%%%%%%%%%%%%%%%%%%%%%%%%%%%%%%%%%%%%%%%%%

\subsubsection{Magnesium cloud structure}
\label{dens_prof}

\begin{figure}[]
\includegraphics[trim=5cm 1cm 6.5cm 7cm, clip=true,width=\columnwidth]{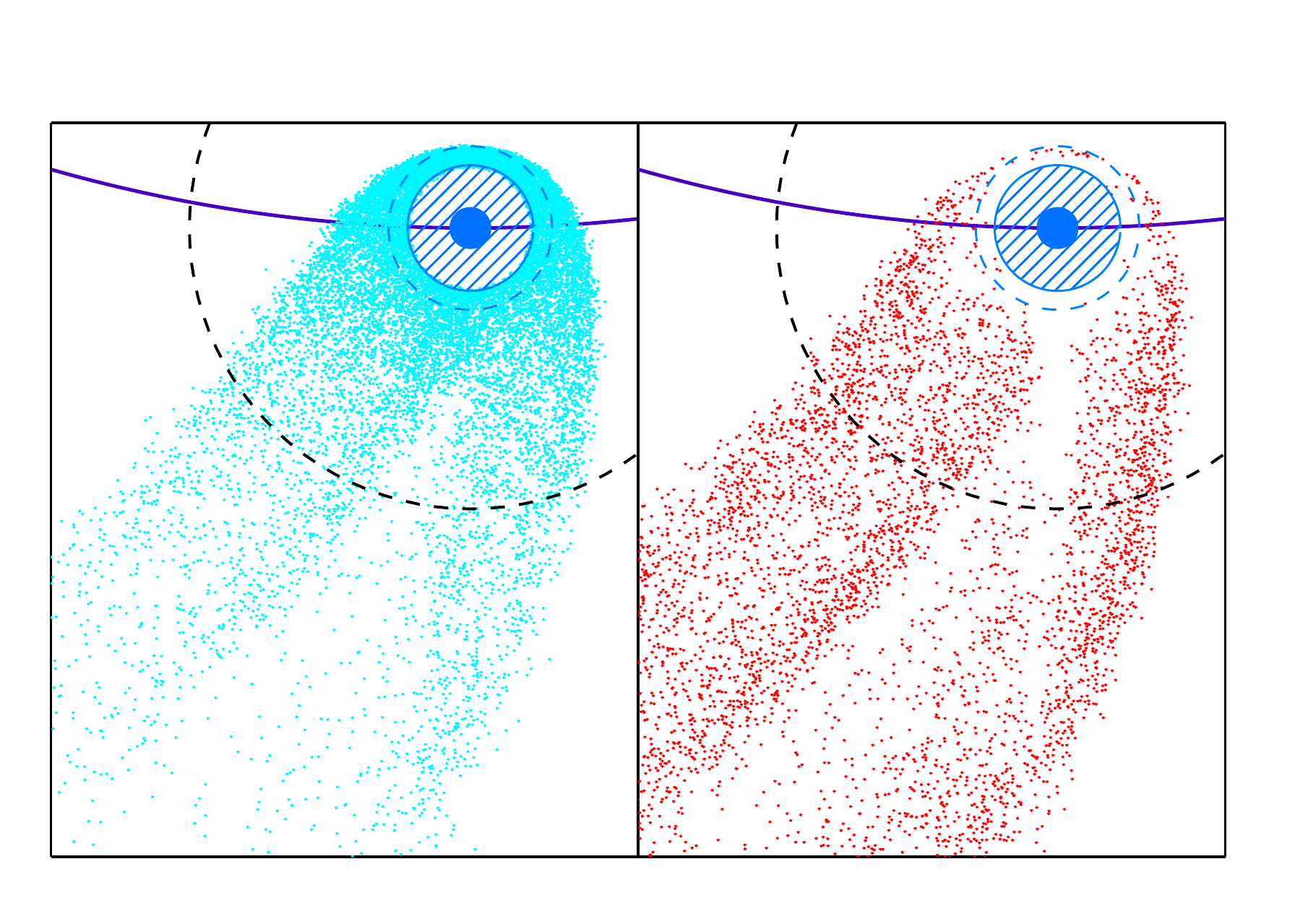}	
\caption[]{Distribution of neutral (light blue dots; left panel) and ionized (light red dots; right panel) magnesium atoms within $\pm$0.5\,$R_{\mathrm{p}}$ of the orbital plane for the best-fit simulation in the G-scenario. Magnesium atoms that escape at the exobase level ($R_{\mathrm{exo}}$=3$\,R_{\mathrm{p}}$; striped blue disk) are accelerated away from the star by radiation pressure (the star is along the vertical direction toward the top of the plot). They remain neutral up to the recombination altitude ($R_{\mathrm{rec}}$=3.9$\,R_{\mathrm{p}}$; blue dashed circle) and keep a high probability to stay in this state until they reach the equilibrium altitude ($R_{\mathrm{eq}}$=13.4$\,R_{\mathrm{p}}$; black dashed circle), beyond which ionized magnesium particles do not recombine efficiently anymore. Because of the planet shadow and shielding from the main atmosphere, the inner part of the cometary tail is empty of radiation-pressure accelerated particles.} 
\label{above_view}
\end{figure}

The extended atmosphere of magnesium in the global best-fit simulation can be seen in Fig.~\ref{above_view}, shaped into a cometary tail by radiation pressure. We measured the mean radial density profiles of neutral and ionized particles inside the cometary tail (Fig.~\ref{dens_prof}) and found that despite their different dynamics the two populations follow Eq.~\ref{pop_ratio}. The magnesium gas is maintained in a nearly full neutral state by electron-recombination up to about 4$\,R_{\mathrm{p}}$, as expected from the value of the recombination altitude (3.9$\,R_{\mathrm{p}}$) associated with the electron density in the global best fit (Fig.~\ref{rate_recAP}). The constant high radial density below $R_{\mathrm{rec}}$ is responsible for the deep theoretical absorption in the core of the Mg\,{\sc i} line. At higher altitudes, the density of neutral magnesium decreases until it is balanced by the density of ionized magnesium near the equilibrium altitude (13.4$\,R_{\mathrm{p}}$). Above the Mg-rec layer there are not enough electrons to sustain the recombination mechanism efficiently and the cloud is mostly ionized. At all altitudes the ionized magnesium population is maintained at low-density levels which, in addition to the low oscillator strengths of the Mg\,{\sc ii}-k and Mg\,{\sc ii}-h lines, ensure that no significant absorption is generated in the Mg\,{\sc ii} doublet (\citealt{VM2013}). Note that because of the asymetry of the escaping cloud the density profiles displayed in Fig.~\ref{dens_prof} may vary on the starward side.\\
The density of neutral magnesium corresponds to typical hydrogen density in the order of $\sim10^{6}\,cm^{-3}$ at the level of the Roche lobe, consistent with the results of \citet{Bourrier_lecav2013}. The resulting mean free path is significantly larger than the atmospheric scale height of HD\,209458b (about 5$\times10^8$cm), and magnesium particles are unlikely to be subjected to collisions above 3$\,R_{\mathrm{p}}$.

\begin{figure}[]
\includegraphics[trim=1cm 1.5cm 2cm 2cm, clip=true,width=\columnwidth]{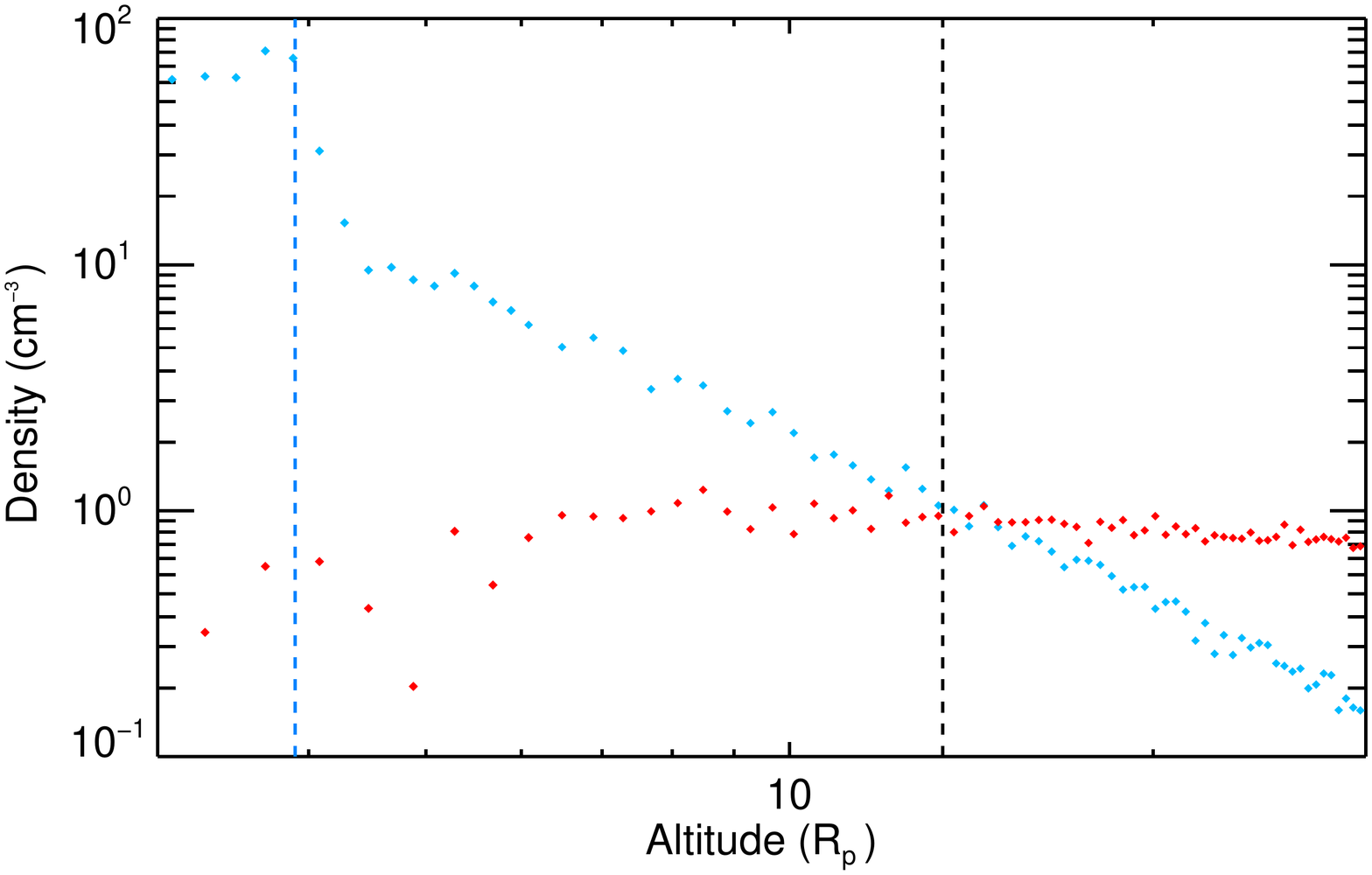} 
\caption[]{Mean radial density profiles of the neutral (blue dots) and singly ionized (red dots) magnesium populations in the cometary tail. Profiles correspond to the best-fit simulation in the G-scenario and are plotted from the exobase level ($R_{\mathrm{exo}}$=3$\,R_{\mathrm{p}}$). Electron-recombination compensates for ionization in the Mg-rec layer below $R_{\mathrm{eq}}$=13.4$\,R_{\mathrm{p}}$ (black dashed line), with a very high efficiency below $R_{\mathrm{rec}}$=3.9$\,R_{\mathrm{p}}$ (blue dashed line). Density of ionized magnesium below $\sim$5\,$R_{\mathrm{p}}$ must be taken with caution as ionized particles are few and far between at these altitudes.} 
\label{dens_prof}
\end{figure}

\subsubsection{Influence of the escape rate and electron density}
\label{effects}

\begin{figure}[]
\includegraphics[trim=1cm 2cm 2cm 2cm, clip=true,width=\columnwidth]{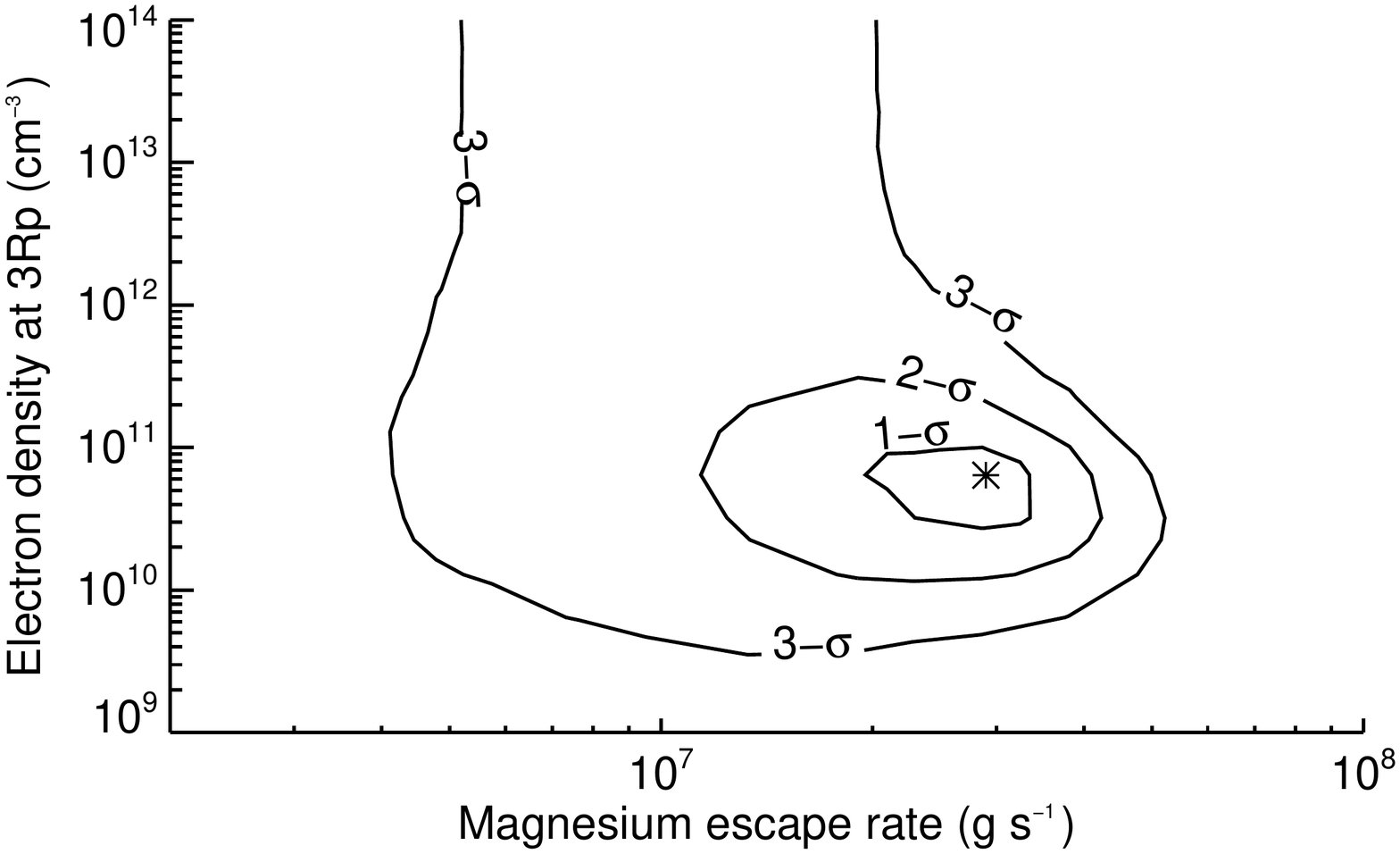} 
\caption[]{Error bars for the estimated neutral magnesium escape rate and electron density at 3\,$R_{\mathrm{p}}$, in the G-scenario. The exobase is fixed at $R_{\mathrm{exo}}=3R_{\mathrm{p}}$ with a planetary wind velocity $v_{\mathrm{pl-wind}}=$25\,km\,s$^{-1}$. A black star indicates the global best fit to the observations with $\dot{M}_{\mathrm{Mg^{0}}}=2.9\times10^{7}$\,g\,s$^{-1}$ and $n_{\mathrm{e}}(3R_{\mathrm{p}})=6.4\times10^{10}$cm$^{-3}$ ($\chi^2$=802.4 for 1067 degrees of freedom).}
\label{bestfitchi2_all}
\end{figure}

\begin{figure}[]
\includegraphics[trim=1cm 2cm 2cm 2cm, clip=true,width=\columnwidth]{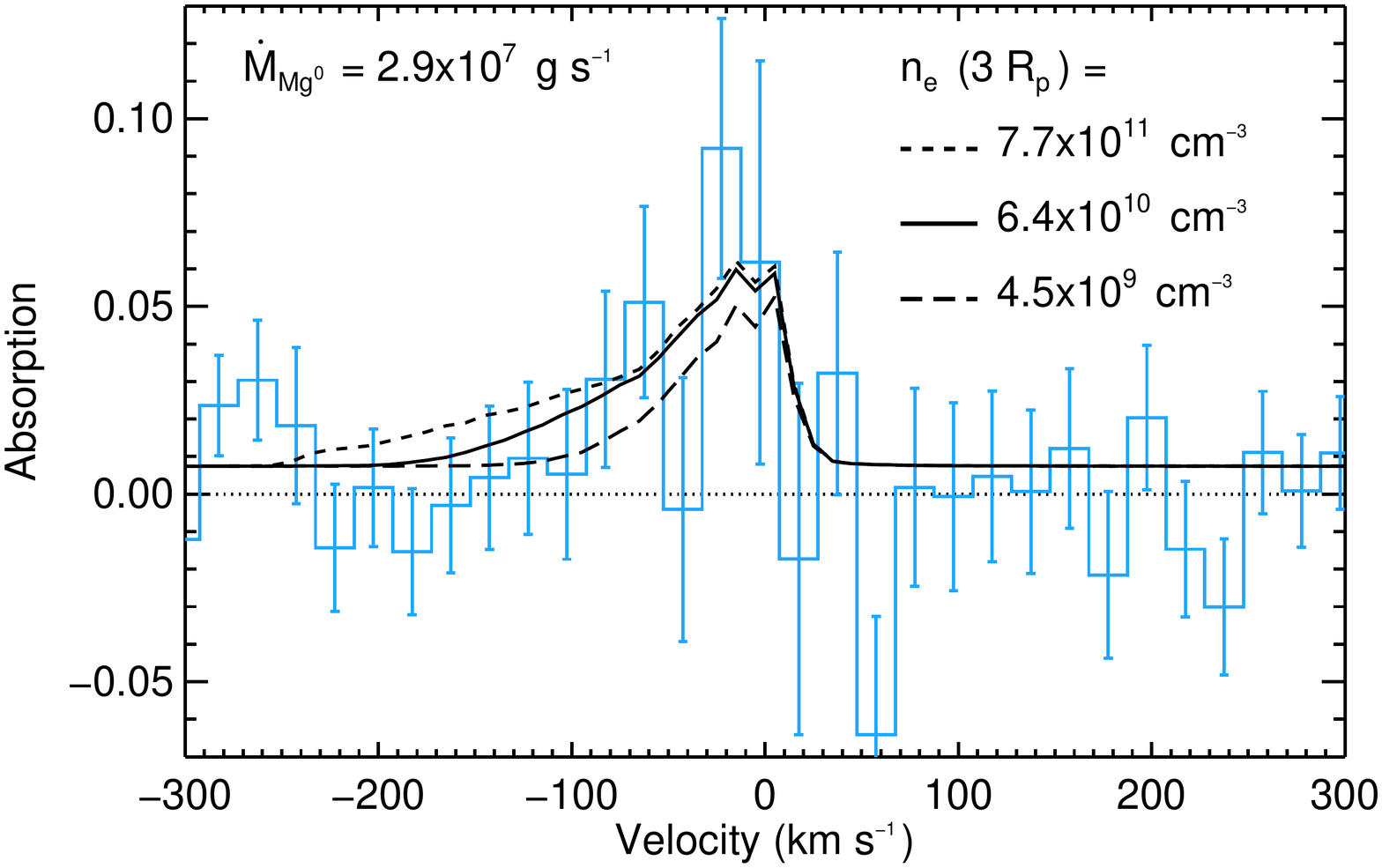}	
\caption[]{Same as in Fig.~\ref{best_absprof}, with two additional theoretical absorption profiles calculated with the same magnesium escape rate $\dot{M}_{\mathrm{Mg^{0}}}=2.9\times10^{7}$\,g\,s$^{-1}$ and electron densities corresponding to $3\sigma$ limits ($\chi^2=$811.4) from the global best fit with $n_{\mathrm{e}}(3R_{\mathrm{p}})=4.5\times10^{9}$cm$^{-3}$ (long-dashed black line) and $n_{\mathrm{e}}(3R_{\mathrm{p}})=7.7\times10^{11}$cm$^{-3}$ (dashed black line). The electron density mostly impacts the width of the absorption profile in its blue wing.} 
\label{tau_fixe_3Rp}
\end{figure}

\begin{figure}[]
\includegraphics[trim=1cm 2cm 2cm 2cm, clip=true,width=\columnwidth]{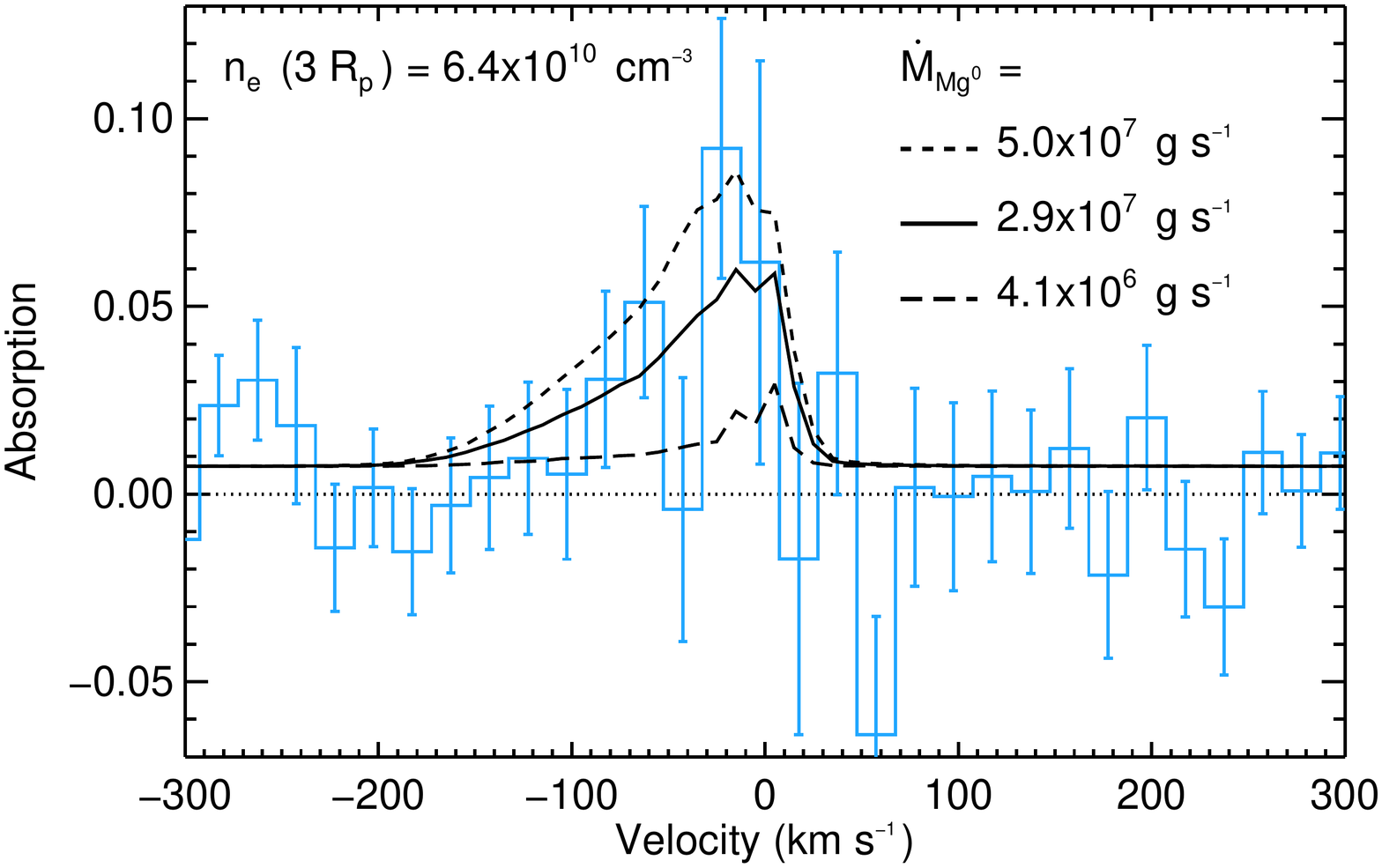}	
\caption[]{Same as in Fig.~\ref{best_absprof}, with two additional theoretical absorption profiles calculated with the same reference electron density $n_{\mathrm{e}}(3R_{\mathrm{p}})=6.4\times10^{10}$cm$^{-3}$ and magnesium escape rates corresponding to $3\sigma$ limits ($\chi^2=$811.4) from the global best fit with $\dot{M}_{\mathrm{Mg^{0}}}=4.1\times10^{6}$\,g\,s$^{-1}$ (long-dashed black line) and $\dot{M}_{\mathrm{Mg^{0}}}=5.0\times10^{7}$\,g\,s$^{-1}$ (dashed black line). The increase in escape rate raises the depth of the absorption profile with little effect on its velocity range.} 
\label{ne_fixe_3Rp}
\end{figure}

We set the exobase properties to the global best-fit values ($R_{\mathrm{exo}}$=$3\,R_{\mathrm{p}}$; $v_{\mathrm{pl-wind}}$=25\,km\,s$^{-1}$) and studied the goodness of fit as a function of the neutral magnesium escape rate and the reference electron density (Fig.~\ref{bestfitchi2_all}). Observations are reproduced up to the 3$\sigma$ level with $n_{\mathrm{e}}(3R_{\mathrm{p}})\ge3.4\times10^{9}$cm$^{-3}$, which corresponds to $R_{\mathrm{eq}}\ge4.9\,R_{\mathrm{p}}$. With these high electron densities, neutral magnesium particles escape the main atmosphere into the Mg-rec layer and have extended lifetimes that lead to absorption at higher velocities in the blue wing of the Mg\,{\sc i} line (Fig.~\ref{tau_fixe_3Rp}). In particular, simulations within $1\sigma$ from the best fit are obtained with reference electron densities in the range $2.7\times10^{10}$ -- $10^{11}$\,cm$^{-3}$ ($R_{\mathrm{eq}}$=8.9 -- 16.5$\,R_{\mathrm{p}}$). In this case, the recombination altitude is also higher than the exobase and all escaping particles stay neutral up to $R_{\mathrm{rec}}$=3.4 -- 4.1$\,R_{\mathrm{p}}$. An increase in the electron density above $\sim$1.2$\times10^{13}$\,cm$^{-3}$ has no influence on the quality of the fit (Fig.~\ref{bestfitchi2_all}) as it corresponds to $R_{\mathrm{rec}}\gtrsim$18$\,R_{\mathrm{p}}$, which is the distance between the star and the planet in the plane of sky at the end of the post-transit observation (\citealt{VM2013}). All the regions of the magnesium cloud observed transiting the stellar disk are thus below $R_{\mathrm{rec}}$ and fully neutral. Alternatively, with lower values of $n_{\mathrm{e}}(3R_{\mathrm{p}})$ ionization losses of neutral magnesium could be compensated for by higher escape rates. However, the subsequent increase in the ionized magnesium population and the resulting absorption in the Mg\,{\sc ii} lines would not be consistent with the observations.\\
We also inferred $1\sigma$ error bars on the escape rate of neutral magnesium in the range $2\times10^{7}$ -- $3.4\times10^{7}$\,g\,s$^{-1}$. In contrast to the reference electron density, the escape rate controls the depth of the absorption profile, but has little import on its velocity range (Fig.~\ref{ne_fixe_3Rp}; see also discussion in \citealt{VM2013}). Best-fit parameters in the G-scenario are similar to those obtained by \citet{VM2013} with $\dot{M}_{\mathrm{Mg^{0}}}=3\times10^{7}$\,g\,s$^{-1}$ and an ionization cut-off radius of 7.5\,$R_{\mathrm{p}}$ located between best-fit values $R_{\mathrm{rec}}$=3.9$\,R_{\mathrm{p}}$ and $R_{\mathrm{eq}}$=13.4$\,R_{\mathrm{p}}$. If we assume a solar abundance (mass fraction Mg/H $\sim9.6\times10^{-4}$; \citealt{Asplund2009}), escape rates of neutral magnesium within $3\sigma$ of our best fit correspond to escape rates of neutral hydrogen in the range $4.3\times10^{9}$ -- $5.4\times10^{10}$\,g\,s$^{-1}$, consistent with the values derived from Lyman-$\alpha$ observations (e.g., \citealt{Ehrenreich2008}; Bourrier \& Lecavelier 2013) and theoretical models (e.g., \citealt{Lecav2004}; \citealt{Yelle2004}; \citealt{Tian2005}). \\

%%%%%%%%%%%%%%%%%%%%%%%%%%%%%%%%%%%%%%%%%%%%%%%%%%%%%%%%%%%%%%%%%%%%%%%%%%%%%%%%%%%%%%%%%%%%%%%%%%%%%%%%%%%%%%%%%%%%%%%%%%%%%%%%%%%%%%%%%%%%%%%%%%%%%%%%%%%%%%%%%%%%%%

\subsubsection{Exobase properties}
\label{exo_prop}

\begin{figure}[]
\includegraphics[trim=2cm 2.5cm 3cm 2.5cm, clip=true,width=\columnwidth]{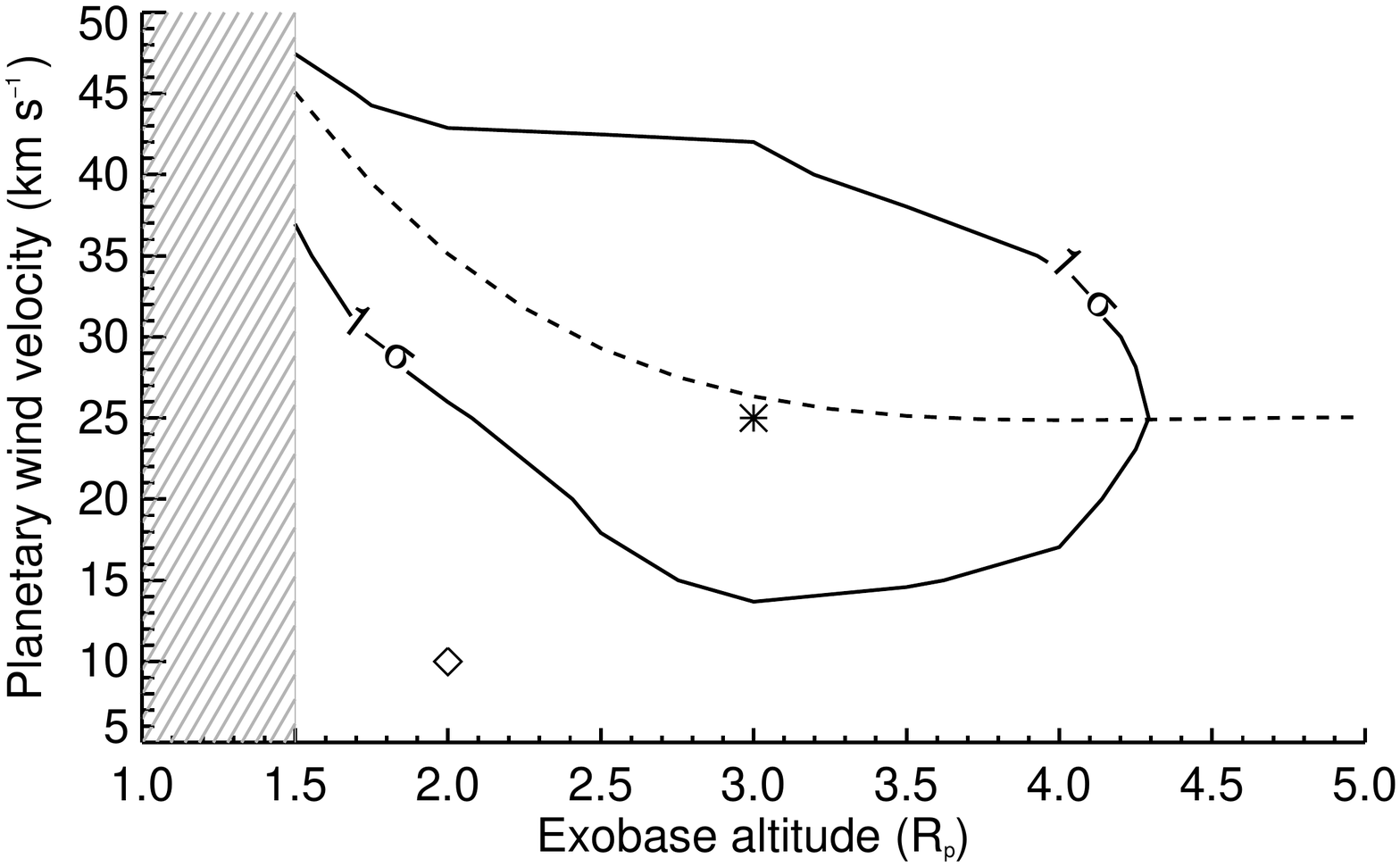}   
\caption[]{1$\sigma$ error bars on the planetary wind velocity and the exobase altitude in the G-scenario. The other model parameters (escape rate and electron density) are also let free. The dashed line shows the fit to the planetary wind velocities that best reproduce the observations for a given exobase altitude. The global best-fit simulation (black star) is obtained with $R_{\mathrm{exo}}=3R_{\mathrm{p}}$, $v_{\mathrm{pl-wind}}=25\,km\,s^{-1}$, $\dot{M}_{\mathrm{Mg^{0}}}$=2.9$\times10^{7}$\,g\,s$^{-1}$ and $n_{\mathrm{e}}(3R_{\mathrm{p}})$=6.4$\times10^{10}$cm$^{-3}$ ($\chi^2_{\mathrm{G-scen}}=802.4$). A black diamond shows the best-fit simulation in the constrained scenario ($v_{\mathrm{pl-wind}}=10\,km\,s^{-1}$; $R_{\mathrm{exo}}=2\,R_{\mathrm{p}}$; $\chi^2_{\mathrm{C-scen}}=807.4$).}
\label{chi2_with_r_v}
\end{figure}

Here we studied the goodness of fit as a function of the exobase altitude and the planetary wind velocity at the exobase, when all four parameters of the model are let free to vary (Fig.~\ref{chi2_with_r_v}). Remarkably, the global best-fit exobase altitude is found to be just above the Roche lobe (about 2.8\,$R_{\mathrm{p}}$ for HD\,209458b) with $R_{\mathrm{exo}}$=3\,$R_{\mathrm{p}}$ and $v_{\mathrm{pl-wind}}$=25\,km\,$s^{-1}$. We looked for the planetary wind velocities that best reproduce the observations for a given exobase altitude. For particles launched beyond the Roche lobe, the planet gravity is negligible compared to stellar gravity and radiation pressure. Consequently, their dynamics only depend on their initial velocity and observations remain best-fitted with $v_{\mathrm{pl-wind}}$=25\,km\,$s^{-1}$. Alternatively, with exobase altitudes below the Roche lobe, planet gravity limits the expansion of the magnesium cloud and the best-fit velocity needs to be higher, up to 45\,km\,$s^{-1}$ at 1.5$\,R_{\mathrm{p}}$. However, as it seems unlikely that the planetary wind is so fast, we fixed $v_{\mathrm{pl-wind}}$=25\,km\,$s^{-1}$ and inferred $1\sigma$ limits on the exobase altitude between 2.1 and 4.3\,$R_{\mathrm{p}}$. Within this range, best fits are always obtained with $\dot{M}_{\mathrm{Mg^{0}}}$=2.9$\times10^{7}$\,g\,s$^{-1}$ and $n_{\mathrm{e}}(3R_{\mathrm{p}})$=6.4$\times10^{10}$cm$^{-3}$ (see Sect.~\ref{tempvar} for more details). \\

Because the atmosphere beyond the Roche lobe is outside of the gravitational influence of the planet it cannot be described accurately with a spherically symmetric hydrostatic profile when the exobase radius exceeds 2.8\,$R_\mathrm{p}$. Besides, gravitational equipotentials near the Roche lobe stray from the spherical approximation used in our model to describe the exobase (see the 3D plot of the shape of the exobase close to the Roche lobe in Fig.~2 of \citealt{Lecav2004}). On the other hand, theoretical models with hydrodynamical escape (\citealt{Yelle2004}; \citealt{GarciaMunoz2007}; \citealt{Koskinen2013a}; \citealt{Guo2013}) predict planetary wind velocities in the range 1 to 10\,km\,s$^{-1}$ at altitudes between 2 and 3\,$R_\mathrm{p}$, which is significantly lower than our global best-fit value of 25\,km\,s$^{-1}$. Therefore, in addition to the G-scenario we considered an alternative constrained scenario (hereafter called C-scenario) in agreement with theoretical predictions for $R_\mathrm{exo}$ and $v_\mathrm{pl-wind}$. In this C-scenario we fixed the exobase at an intermediate altitude of 2\,$R_\mathrm{p}$ and constrained the planetary wind to 10\,km\,s$^{-1}$ (Fig.~\ref{chi2_with_r_v}). We study the atmospheric escape in this scenario in the following section.

%%%%%%%%%%%%%%%%%%%%%%%%%%%%%%%%%%%%%%%%%%%%%%%%%%%%%%%%%%%%%%%%%%%%%%%%%%%%%%%%%%%%%%%%%%%%%%%%%%%%%%%%%%%%%%%%%%%%%%%%%%%%%%%%%%%%%%%%%%%%%%%%%%%%%%%%%%%%%%%%%%%%%%

\subsection{Constrained scenario}
\label{cons_approach}

\subsubsection{Physical conditions in HD\,209458b exosphere}

\begin{figure}[]
\includegraphics[trim=1cm 2cm 2cm 2cm, clip=true,width=\columnwidth]{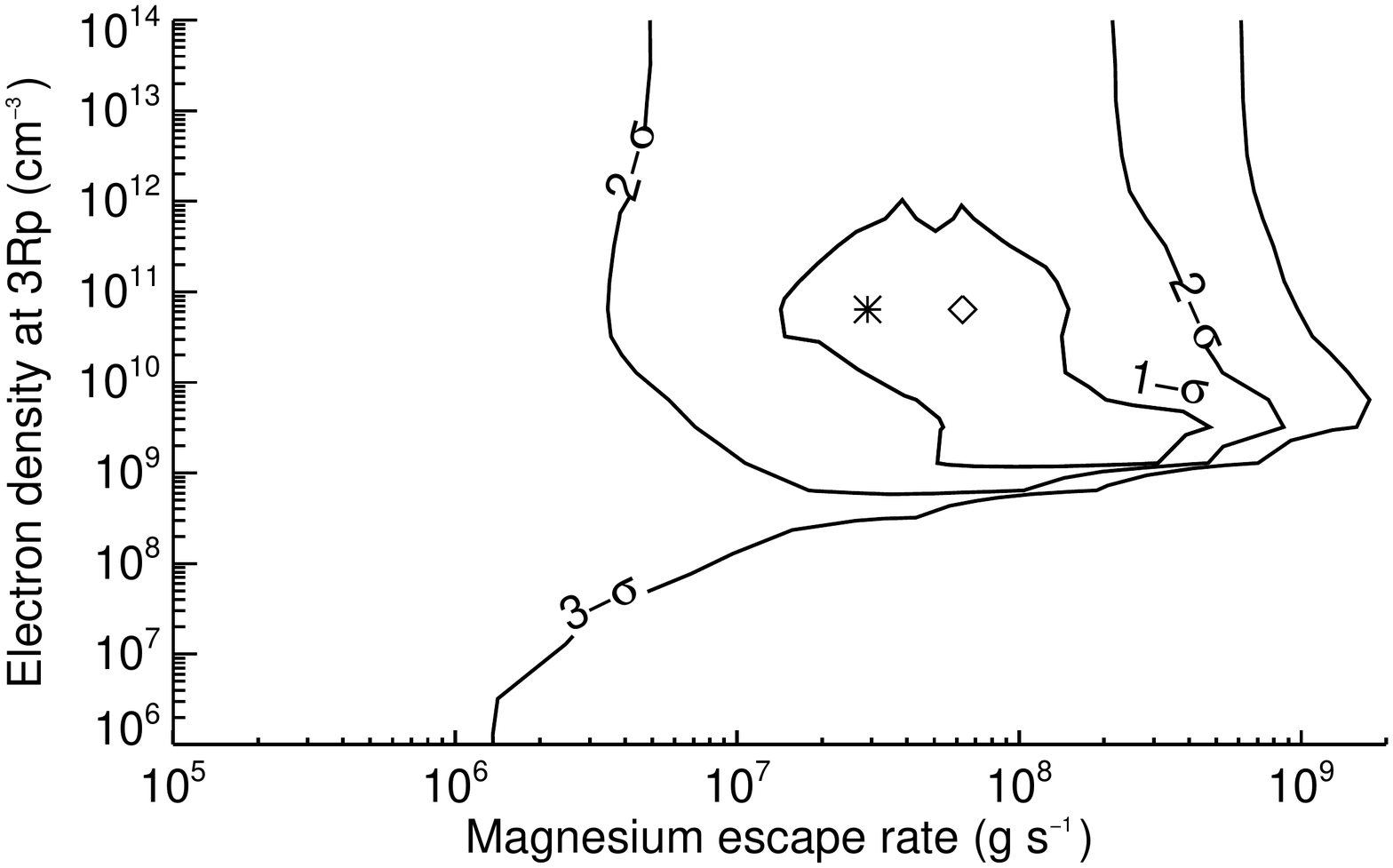}		
\caption[]{Error bars for the estimated magnesium escape rate and electron density at 3\,$R_{\mathrm{p}}$, in the C-scenario. The exobase is fixed at $R_{\mathrm{exo}}=2R_{\mathrm{p}}$ with a planetary wind velocity $v_{\mathrm{pl-wind}}=$10\,km\,s$^{-1}$. A diamond indicates the constrained best fit to the observations with $\dot{M}_{\mathrm{Mg^{0}}}=6.3\times10^{7}$\,g\,s$^{-1}$ ($\chi^2$=807.4 for 1069 degrees of freedom), while a star shows the global best fit with $\dot{M}_{\mathrm{Mg^{0}}}=2.9\times10^{7}$\,g\,s$^{-1}$ ($\chi^2$=802.4 for 1067 degrees of freedom), in both cases obtained with $n_{\mathrm{e}}(3R_{\mathrm{p}})=6.4\times10^{10}$cm$^{-3}$.}
\label{bestfitchi2}
\end{figure}

In the C-scenario ($v_{\mathrm{pl-wind}}=10\,km\,s^{-1}$; $R_{\mathrm{exo}}=2\,R_{\mathrm{p}}$) the \textit{constrained} best fit to the observations is found with a $\chi^2$ of 807.4 for 1069 degrees of freedom (Fig.~\ref{bestfitchi2}), at little more than 2$\sigma$ from the global best fit, and yields similar values for the magnesium escape rate and reference electron density ($\dot{M}_{\mathrm{Mg^{0}}}$=6.3$\times10^{7}$\,g\,s$^{-1}$; $n_{\mathrm{e}}(3R_{\mathrm{p}})$=6.4$\times10^{10}$cm$^{-3}$) albeit with larger error bars, in particular at the $1\sigma$ level ($1.4\times10^{7}$ -- $4.7\times10^{8}$\,g\,s$^{-1}$ and $1.2\times10^{9}$ -- $9\times10^{11}$cm$^{-3}$). As in the G-scenario there is no change in the fit quality (here within 2$\sigma$ from the best fit) with electron density above $\sim$1.2$\times10^{13}$\,cm$^{-3}$. On the contrary, the fit becomes worse with escape rates in the range 5$\times10^{7}$ -- 5$\times10^{8}$\,g\,s$^{-1}$ and $n_{\mathrm{e}}(3R_{\mathrm{p}})\lesssim10^{9}$cm$^{-3}$ because the recombination altitude then falls below the exobase. A significant portion of the escaping gas is ionized and, as the atmosphere is denser in the C-scenario, Mg\,{\sc ii} line absorption depths increase steeply with decreasing reference electron densities. Assuming electrons are mainly produced by the ionization of hydrogen most theoretical models of HD\,209458b atmosphere predict electron densities in the range $10^{6}$ -- $10^{7}$cm$^{-3}$ at 3\,$R_{\mathrm{p}}$ (e.g., \citealt{Yelle2004}; \citealt{Koskinen2013a}; \citealt{Guo2013}). In the C-scenario this situation corresponds to a regime with no recombination above the exobase and can only reproduce the observations within 3$\sigma$ from the constrained best fit, with escape rates below $\sim$2$\times10^{6}$\,g\,s$^{-1}$. In this case, absorption profiles in the magnesium lines are shallow and not much influenced by variations of $n_{\mathrm{e}}(3R_{\mathrm{p}})$ and $\dot{M}_{\mathrm{Mg^{0}}}$ (Fig.~\ref{bestfitchi2}). \\

%%%%%%%%%%%%%%%%%%%%%%%%%%%%%%%%%%%%%%%%%%%%%%%%%%%%%%%%%%%%%%%%%%%%%%%%%%%%%%%%%%%%%%%%%%%%%%%%%%%%%%%%%%%%%%%%%%%%%%%%%%%%%%%%%%%%%%%%%%%%%%%%%%%%%%%%%%%%%%%%%%%%%%%%

\subsubsection{Radiation pressure and planetary gravity}
\label{gravSS}

\begin{figure*}
\centering
\begin{minipage}[b]{0.8\textwidth}	
\includegraphics[width=\columnwidth]{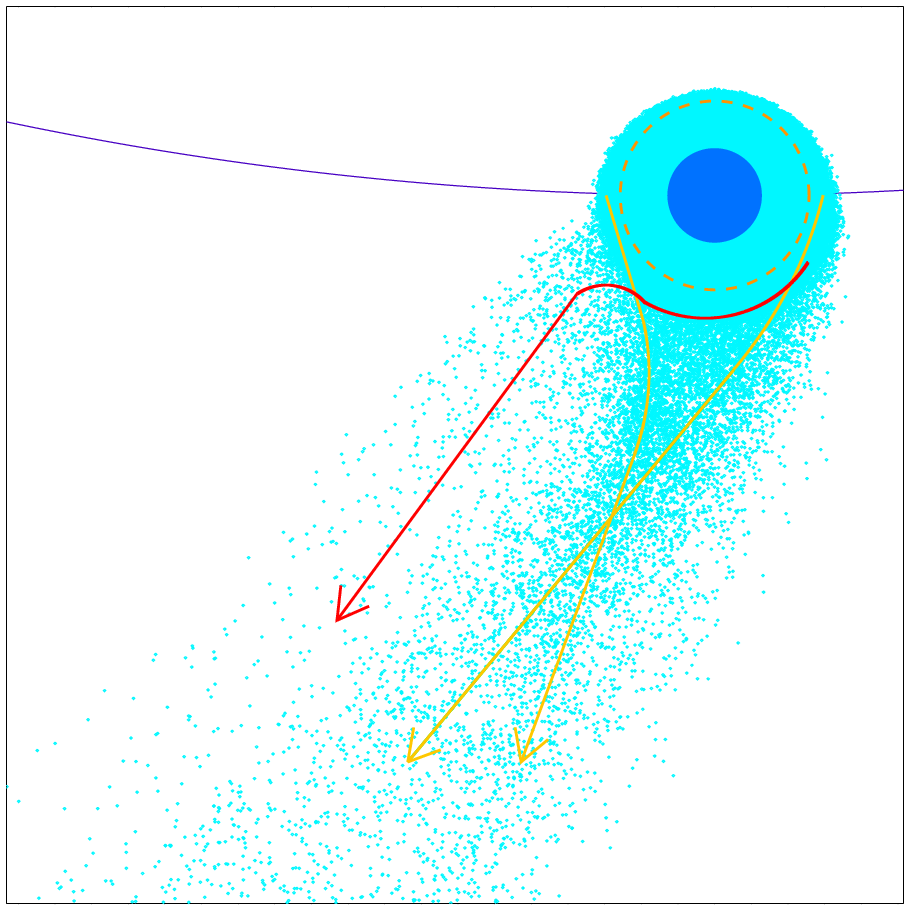}
\caption[]{View of the neutral magnesium cloud from above of the orbital plane in the best-fit simulation (C-scenario). Because of the high density below the exobase ($R_{\mathrm{exo}}$=2$\,R_{\mathrm{p}}$; orange dashed circle) magnesium atoms escape more easily from the limbs of the atmosphere, unshielded from radiation pressure (the star is toward the top of the plot). The orange lines show the average trajectory of these flows. Gravitational effects and the lower velocity of the planetary wind prevent the cometary tail from dispersing as in the G-scenario (Fig.~\ref{above_view}). The red line shows the average trajectory of the flow of particles emanating from shielded regions. Deviated by planetary gravity, the flow emerges from the cloud to the rear of the planet, where it becomes subjected to radiation pressure.} 
\label{orb_plane}
\end{minipage}
\end{figure*}

In the case of the global best fit ($R_{\mathrm{exo}}=3\,R_{\mathrm{p}}$ and $v_{\mathrm{pl-wind}}=$25\,km\,s$^{-1}$), the gas cloud is mostly optically thin and spread over a large area. By contrast when particles are launched with lower velocities at altitudes below the Roche lobe, as for the constrained best fit ($R_{\mathrm{exo}}=2\,R_{\mathrm{p}}$ and $v_{\mathrm{pl-wind}}=$10\,km\,s$^{-1}$), densities of neutral magnesium are high enough for the bulk of the atmosphere to be self-shielded from stellar photons and particles escape more easily from the unshielded limbs of the atmosphere (Fig.~\ref{orb_plane}). Besides, at the low altitude of the exobase in the constrained best fit, planetary gravity is the dominant force and it strongly influences the dynamics of escaping particles. Those escaping from shielded regions are so deflected that they move backward of the planet orbital motion until they emerge from the other side of the cometary tail to the rear of the planet (Fig.~\ref{orb_plane}).\\

%%%%%%%%%%%%%%%%%%%%%%%%%%%%%%%%%%%%%%%%%%%%%%%%%%%%%%%%%%%%%%%%%%%%%%%%%%%%%%%%%%%%%%%%%%%%%%%%%%%%%%%%%%%%%%%%%%%%%%%%%%%%%%%%%%%%%%%%%%%%%%%%%%%%%%%%%%%%%%%%%%%%%%%%%%%%%%%%

\subsection{Influence of the main atmosphere temperature}
\label{main_atm_temp}

\begin{figure}[]
\includegraphics[trim=2cm 2cm 2cm 2cm, clip=true,width=\columnwidth]{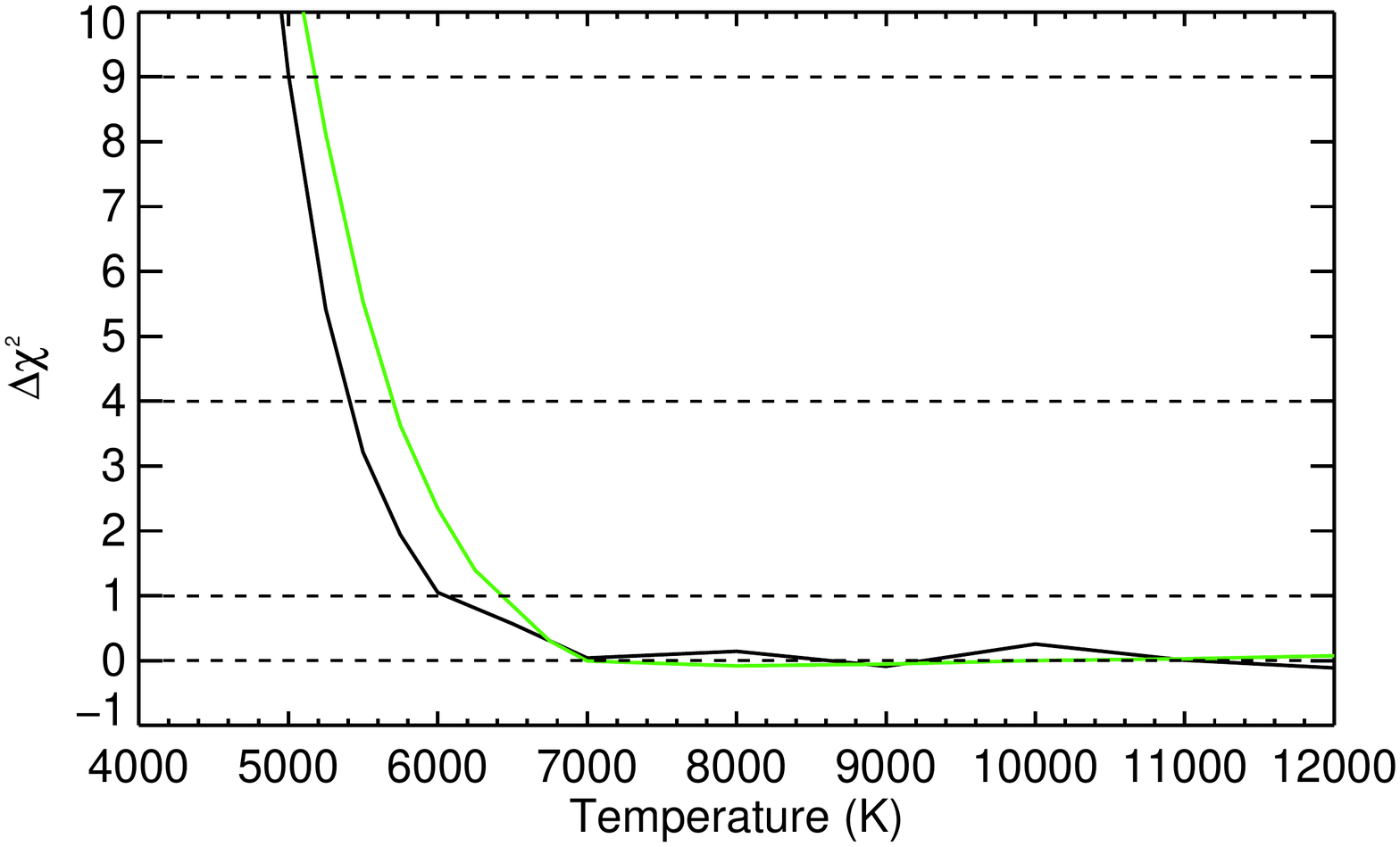}	
\caption[]{Plot of the $\chi^2$ difference between simulations obtained with the best-fit parameters of the G-scenario (black line) or the C-scenario (green line) as a function of the main atmosphere temperature. Horizontal black dotted lines are plotted at 1, 2, and 3$\sigma$ from each best fit ($\chi^2_{\mathrm{G-scen}}$=802.4; $\chi^2_{\mathrm{C-scen}}$=807.4), and in both cases the $\chi^2$ difference exceeds 1$\sigma$ with temperatures below $\sim6100$\,K.} 
\label{best_chi2_with_T}
\end{figure}

\begin{figure}[]
\includegraphics[trim=1cm 2cm 2cm 2cm, clip=true,width=\columnwidth]{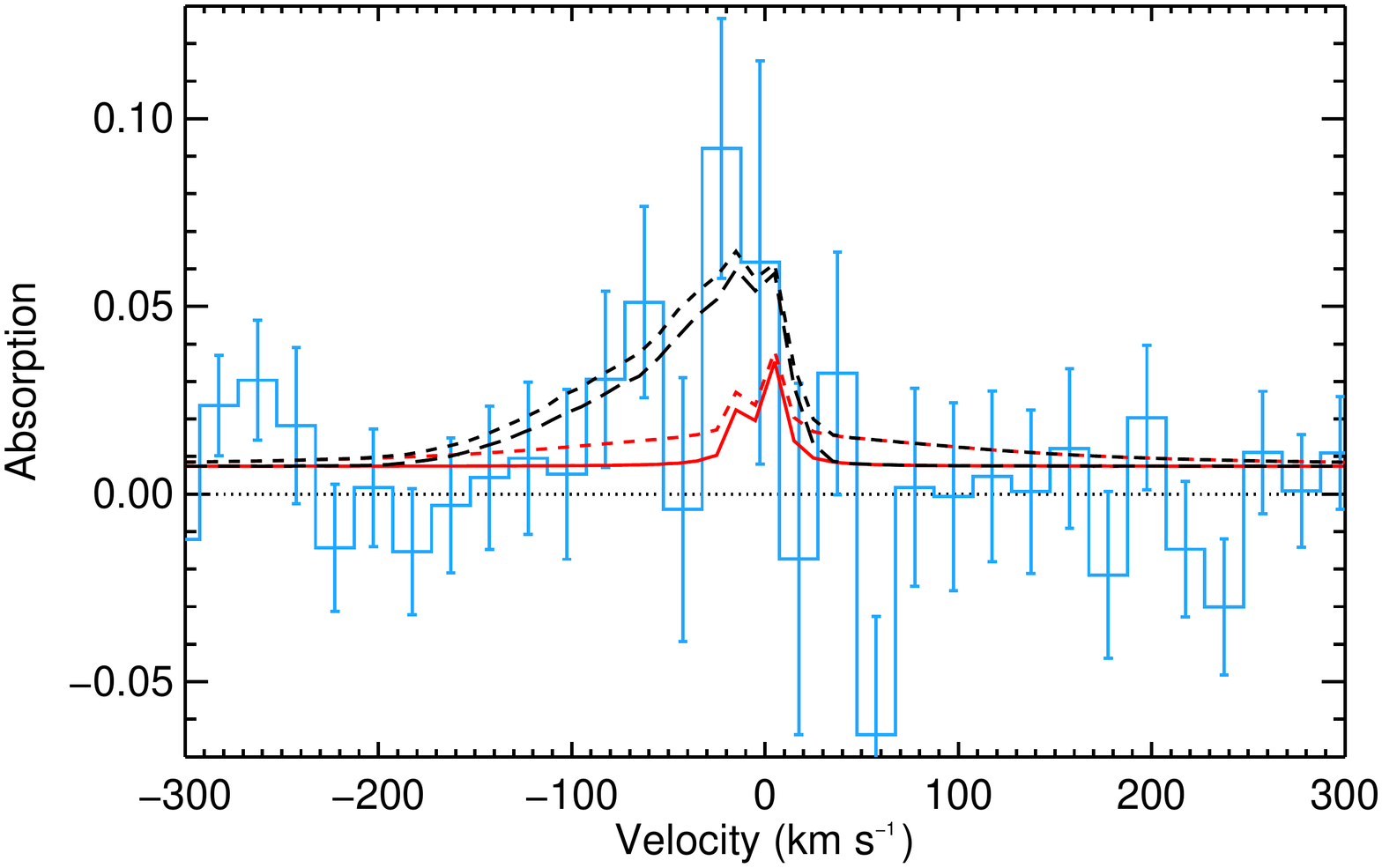}		 
\caption[]{Absorption profile observed in the Mg\,{\sc i} line (blue histogram). Two absorption profiles are calculated for best-fit parameters (G-scenario) and a main atmosphere temperature of $5000\,K$ (dashed black line) and $7000\,K$ (solid black line). In both cases, the red line shows the part of the absorption profile generated by neutral magnesium in the main atmosphere. The decrease in temperature leads to higher densities in the main atmosphere, which increases absorption in the wings of the line through natural broadening.}
\label{abs_T}
\end{figure}

Here we tested the impact of the main atmosphere temperature on the theoretical Mg\,{\sc i} line absorption profile and the fit to the data. We considered no change in the escape rate, the reference electron density and the exobase properties, with the same values that provided the best fit to the observations in either the G-scenario or the C-scenario. We found that an increase in the temperature of the main atmosphere $\overline{T}$ has no influence on the quality of the fits (Fig.~\ref{best_chi2_with_T}). On the other hand, the $\chi^2$ quickly increases as temperature decreases, and we put a 1$\sigma$ lower limit on $\overline{T}$ at about $\sim6100$\,K. If the temperature is lower than this value, densities in the main atmosphere are indeed much higher and result in higher absorption depths in the damped Mg\,{\sc i} line wings, which are not observed (Fig.~\ref{abs_T}).

\subsection{Discussion on the electron density}
\label{ne_discuss}

We obtained electron densities at $3\,R_\mathrm{p}$ significantly higher than the $10^{6}$ -- $10^{7}$cm$^{-3}$ range expected from theoretical models (e.g., \citealt{Guo2013}). It is possible that these values have a large uncertainty, as such models are by necessity based on many assumptions regarding the energy input, the atmospheric processes, and composition, and are furthermore simplified 1D representations of the sharp atmospheric transition between the day and night sides (i.e., along the terminator as observed during transits). If ions and electron densities are indeed similar, collisions of magnesium with ions should modify the dynamics of the escaping gas. However, the hydrogen escape rate, planetary wind velocity and exobase altitude derived from our best fits are consistent with values found in the literature, and are independent of the electron density in their effect on the theoretical absorption profile (see for example Fig.~\ref{tau_fixe_3Rp} and \ref{ne_fixe_3Rp}). This  gives us confidence in our modeling and the assumption that the atmosphere beyond the Roche lobe is collisionless. As a first explanation it is conceivable that electrons may be more abundant than ions in the upper atmosphere. Alternatively, we may have overestimated the electron density. First, we would not detect an absorption signature in the Mg\,{\sc ii} line if the ionized magnesium cloud produced by the atmospheric escape was extending far away from the planet to occult the stellar disk at all observed orbital phases, a scenario which was proposed for WASP-12 b by \citet{Haswell2012}. In this case, the density of neutral magnesium needed to reproduce the observations would be obtained with much lower recombination rates and electron densities because of the large amount of ionized magnesium available for recombination. Second, depending on the heating efficiency and stellar irradiation, temperatures in the extended atmosphere of HD209458b could be significantly higher than the average temperature profile we used (e.g., \citealt{Koskinen2013a}). With temperatures of about 13000\,K, which would be consistent with the high heating efficiency expected in the atmosphere of HD209458b (\citealt{Ehrenreich_desert2011}; \citealt{Koskinen2013a}), best-fit electron densities are in the order of $10^{8}$cm$^{-3}$ because of the higher recombination rate. Third, the radiative and dielectronic recombination rates estimated by \citet{Aldrovandi1973} may be inadequate in the hydrodynamic atmosphere of HD209458b.

%%%%%%%%%%%%%%%%%%%%%%%%%%%%%%%%%%%%%%%%%%%%%%%%%%%%%%%%%%%%%%%%%%%%%%%%%%%%%%%%%%%%%%%%%%%%%%%%%%%%%%%%%%%%%%%%%%%%%%%%%%%%%%%%%%%%%%%%%%%%

\section{Spectro-temporal variability}
\label{tempvar}

\begin{figure*}
\centering
\begin{minipage}[b]{0.9\textwidth}	
\includegraphics[trim=1cm 3cm 2cm 2cm, clip=true,width=\columnwidth]{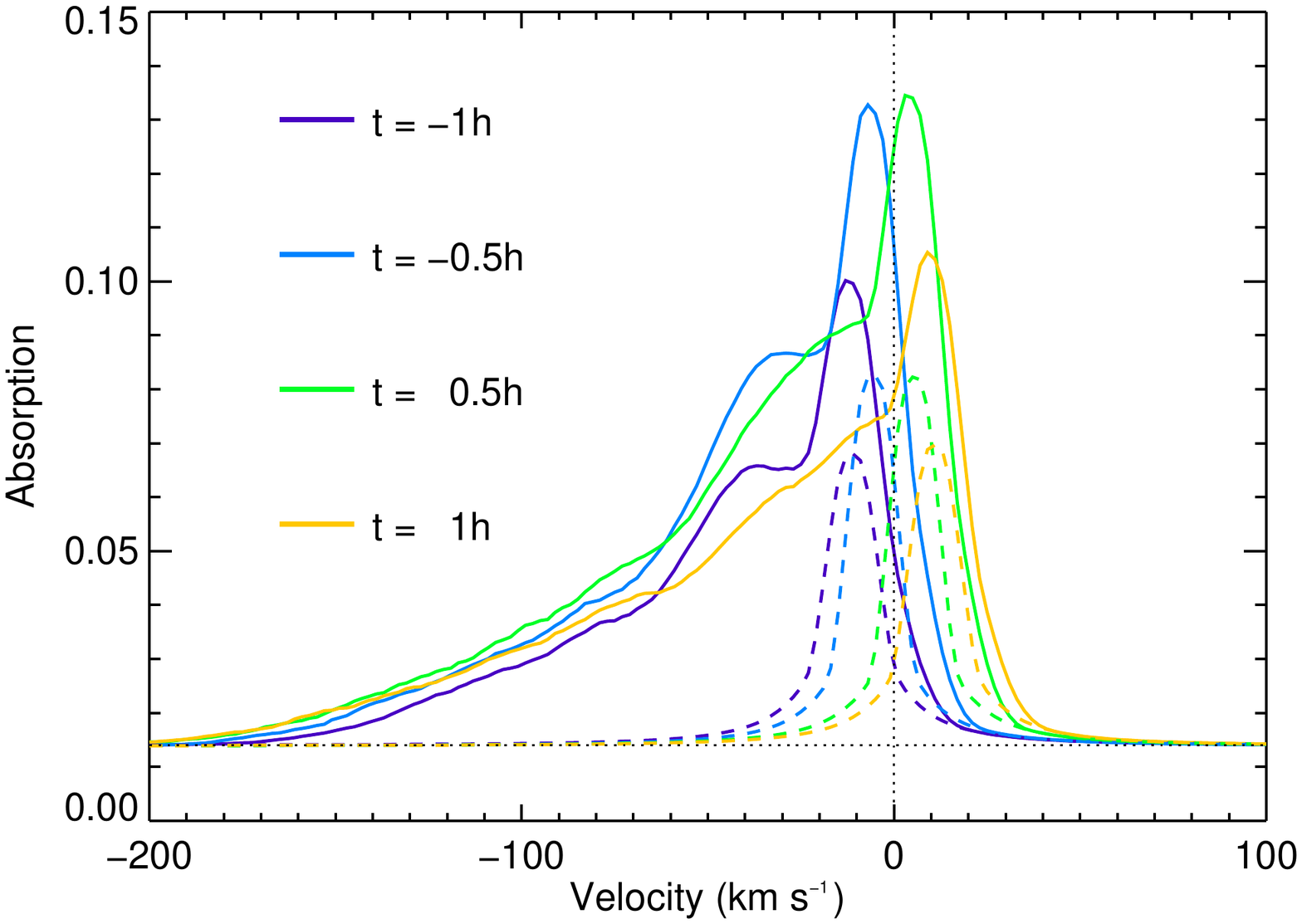}		%ok nlle fig, pour R=3/v=25, smooth 6 (sinon effet de resolution de la thermo donne pic au centre)
\caption[]{Evolution of the theoretical absorption profile as a function of time during the transit of HD\,209458b, for the best-fit simulation of the G-scenario. Absorption in the blue side is generated by radiation-pressure accelerated magnesium atoms and always drops at about -60\,km\,s$^{-1}$ (at mid-height). The central absorption peak comes from atoms close to the planet (dotted lines show the part of the absorption profile generated by the main atmosphere) and shifts redward of the line with the orbital motion. This is similar to the spectral extension of the observed absorption signature toward positive velocities during and after the transit (Fig.~\ref{abs_3obs}). The simulated planetary disk is the source for an absorption depth of $\sim1.4\%$ at all wavelengths during the transit.} 
\label{evol_temp}
\end{minipage}
\end{figure*}

\begin{figure*}
\centering
\begin{minipage}[b]{0.9\textwidth}		
\includegraphics[width=\columnwidth]{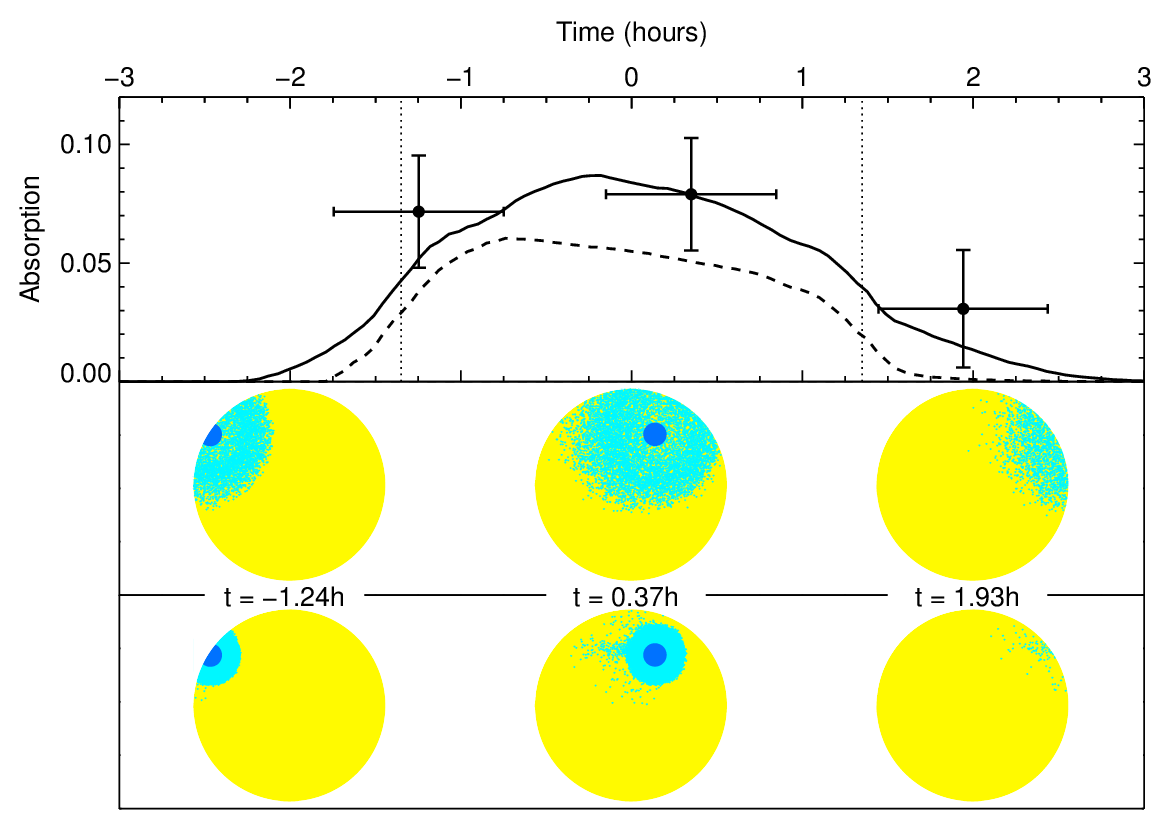}	
\caption[]{\textit{Top panel:} Theoretical transit absorption depth over the range -60 to 0\,km\,s$^{-1}$ in the Mg\,{\sc i} line (solid and dashed lines). Measurements for the transit-ingress, transit-center, and post-transit observations (\citealt{VM2013}) are plotted as black dots with horizontal error bars showing their duration. Vertical dotted lines show the beginning and end of ingress and egress of the planetary transit. The theoretical absorption depth, determined by the stellar surface occulted by the magnesium cloud and its opacity, is displayed for the global best-fit simulation (solid line; $R_{\mathrm{exo}}=3\,R_{\mathrm{p}}$, $v_{\mathrm{pl-wind}}=$25\,km\,s$^{-1}$, $\dot{M}_{\mathrm{Mg}}=2.9\times10^{7}$\,g\,s$^{-1}$) and the constrained best-fit simulation (dashed line; $R_{\mathrm{exo}}=2\,R_{\mathrm{p}}$, $v_{\mathrm{pl-wind}}=$10\,km\,s$^{-1}$,  $\dot{M}_{\mathrm{Mg}}=6.3\times10^{7}$\,g\,s$^{-1}$). \textit{Lower panels:} Views of the magnesium particles with velocities in the range -60 to 0\,km\,s$^{-1}$ along the star/Earth line of sight (light blue). HD\,209458b is displayed as a deep blue disk. Although the cloud occults a larger area in the global best-fit simulation (middle panel), it is optically thinner than in the constrained best-fit simulation (bottom panel).} 
\label{sim_tot_abs_HD209}
\end{minipage}
\end{figure*}

We obtained variations with time in the velocity structure of the best-fit theoretical absorption profile, regardless of the considered scenario (Fig~\ref{evol_temp}). Radiation-pressure-accelerated neutral atoms free of the planet influence absorb in the blue wing of the Mg\,{\sc i} line up to a velocity determined by their lifetime, i.e., by the reference electron density. By contrast radiation pressure from the Mg\,{\sc i} line has little influence on particles close to the planet (which are neutral in the global and constrained best fits), as they have low velocities and are mostly shielded from stellar photons. These particles follow the planet orbital motion and their absorption in the Mg\,{\sc i} line core shifts toward higher positive velocities during the transit, following the increase in the planet velocity on the line of sight between -15 to 15\,km\,s$^{-1}$ from ingress to egress (as observed in the line of carbon monoxide by \citealt{Snellen2010}). This variation is similar to the putative redward extension of the absorption signature reported by \citet{VM2013} between the transit and post-transit observation, as can be seen in Fig.~\ref{abs_3obs}. The ratio between the populations generating abssorption in either the core or the wing of the spectral Mg\,{\sc i} line is constrained by the planetary wind velocity at the exobase, and the best-fit values described in Sect.~\ref{exo_prop} are those that balance the two populations and their respective absorptions in a way consistent with the observations (Fig~\ref{evol_temp}). Independent of the planetary wind velocity and the exobase altitude, the escape rate and reference electron density control the total amount of neutral magnesium particles in the cloud and their maximum velocities, i.e., the overall depth of the absorption profile and its width. As a result, observations are always best reproduced with $n_{\mathrm{e}}(3\,R_\mathrm{p})=6.4\times10^{10}$\,cm$^{-3}$ and $\dot{M}_{\mathrm{Mg^{0}}}$=2.9$\times10^{7}$\,g\,s$^{-1}$, regardless of the best couples ($v_{\mathrm{pl-wind}}$;$R_{\mathrm{exo}}$)  in Fig.~\ref{chi2_with_r_v}. \\
The velocity range of the observed absorption signature also limits the size of the cometary tail of neutral magnesium. For both best-fit simulations it absorbs efficiently up to about 15 planetary radii (Fig.~\ref{above_view}), and seen from the Earth, the projection of the cloud onto the stellar disk extends to only a few planetary radii. Absorption drops quickly after the end of the planetary transit (Fig~\ref{sim_tot_abs_HD209}) and post-transit absorption depths are lower than the value measured by \citet{VM2013}.

%%%%%%%%%%%%%%%%%%%%%%%%%%%%%%%%%%%%

\section{Conclusion}
\label{conclu}

We used a revised version of the 3D particle model described in Bourrier \& Lecavelier (2013) to simulate the escape of magnesium from the atmosphere of HD\,209458b and to calculate theoretical transmission spectra to be compared to transit and post-transit observations obtained in the Mg\,{\sc i} and Mg\,{\sc ii} lines in the UV by \citet{VM2013}. The main improvement to the previous model is a detailed analytical modeling of the atmosphere below the exobase which takes its shielding effect and its absorption in the core of the neutral magnesium line into account. We found that the mean temperature below the exobase must be higher than $\sim6100$\,K. The free parameters of the model are otherwise the escape rate of neutral magnesium, the electron density, the altitude of the exobase, and the velocity of the planetary wind at the exobase. \\
Observations are best explained if the exobase is close to the Roche lobe (\mbox{$R_\mathrm{exo}$=3$\stackrel{+1.3}{_{-0.9}}$$\,R_\mathrm{p}$}) and magnesium atoms escape the atmosphere at the exobase level with a radial velocity $v_{\mathrm{pl-wind}}$=25\,km\,s$^{-1}$. Because of the V-shape of the stellar Mg\,{\sc i} line, radiation pressure is not too efficient on low-velocity neutral magnesium atoms close to the planet but impart strong accelerations to atoms with high radial velocities, as they receive more UV flux far from the line core. The velocity range of the absorption signature, up to -60\,km\,s$^{-1}$ in the blue wing of the line, is thus well reproduced by a radiative blow-out, provided the quick stellar UV-photoionization of the escaping particles is compensated for by electron recombination up to an equilibrium altitude between the two mechanisms, which is estimated to be $R_{\mathrm{eq}}$=13.4$\stackrel{+3.1}{_{-4.5}}$$\,R_{\mathrm{p}}$. Beyond this Mg-recombining layer of the exosphere the magnesium is mostly ionized and subjected to a low radiation pressure from the Mg\,{\sc ii} line. The non-detection of excess absorption in this line is well-explained by low densities of ionized magnesium at all altitudes. The global best fit to the observations is obtained with an escape rate $\dot{M}_{\mathrm{Mg}}=2.9\times10^{7}$\,g\,s$^{-1}$ ($2.0\times10^{7}$ -- $3.4\times10^{7}$\,g\,s$^{-1}$). Assuming a solar abundance these results are consistent with standard values of hydrogen escape rates in the range $2.1\times10^{10}$ -- $3.5\times10^{10}$\,g\,s$^{-1}$. Observations are best reproduced with electron density at $3R_{\mathrm{p}}$ in the order of 10$^{10}$cm$^{-3}$. This value is possibly overestimated; or if the electron density is correct, it could mean that electrons are more abundant than ions in this collisionless part of the upper atmosphere. With the addition of electron-impact ionization to UV-photoionization, electron-recombination is dominated by ionization close to the planet (below 2.7$R_{\mathrm{p}}$), but this has little influence on the quality of the best fit since particles are launched above the Roche lobe. In any case, the rates used by \citet{Voronov1997} for electron-impact ionization may not be valid for temperatures below 11000\,K because they were calculated for plasmas with far higher temperatures. \\
Constraining the exobase radius and planetary wind velocity to lower values consistent with theoretical model predictions ($R_\mathrm{exo}$=2$\,R_\mathrm{p}$ and $v_{\mathrm{pl-wind}}$=10\,km\,s$^{-1}$) we found that the observations can still be explained with similar escape rates and electron densities, albeit with larger error bars: $\dot{M}_{\mathrm{Mg}}$=6.3$\times10^{7}$\,g\,s$^{-1}$ ($1.4\times10^{7}$ -- $4.7\times10^{8}$\,g\,s$^{-1}$). Although in this case the exobase radius was chosen arbitrarily, halfway between the planet surface and the Roche lobe, a different value in this range does not significantly change our conclusions.\\
Simulations show that the absorption profile results from complex interactions between radiation pressure, planetary gravity, and self-shielding. In a previous work, the analysis of transit observations in the Lyman-$\alpha$ line of neutral hydrogen allowed us to characterize atmospheric escape at high altitudes in the exosphere. Here the absorption signature in the line of neutral magnesium has been used to analyze the structure of the atmosphere at lower altitudes both in the escaping cloud and below the exobase. Simulations predict that the absorption profile should expand toward positive velocities during the transit because the atmosphere close to the planet follows its circular orbital motion. Observation of the magnesium lines thus appears to be a powerful tool to probe an exoplanet's upper atmosphere in the thermosphere-exosphere transition region. We can anticipate that several host stars of transiting planets may be bright enough in the Mg\,{\sc i} line for their atmosphere to be probed in such a way. 

%%%%%%%%%%%%%%%%%%%%%%%%%%%%%%%%%%%%%%%%%%%%%%%%%%%%%%%%%%%%%%%%%%%%%%%%%%%%%%%%%%%%%%%%%%%%%%%%%%%%%%%%%%%%%%%%%%%%%%%%%%

\begin{acknowledgements}
The authors wish to thank the referee Tian Feng for his perceptive comments, which greatly improved this paper and helped to put its results in perspective. Based on observations made with the NASA/ESA Hubble Space Telescope, obtained at the Space Telescope Science Institute, which is operated by the Association of Universities for Research in Astronomy, Inc., under NASA contract NAS 5-26555. The authors acknowledge financial support from the Centre National d'\'Etudes Spatiales (CNES). The authors acknowledge the support of the French Agence Nationale de la Recherche (ANR), under program ANR-12-BS05-0012 ``Exo-Atmos''. This work has also been supported by an award from the Fondation Simone et Cino Del Duca. 
\end{acknowledgements}

\bibliographystyle{aa} % style aa.bst
\bibliography{biblio} % your references Yourfile.bib

\end{document}